# Chongqing University of Posts and Telecommunications

# Thesis for Master's Degree

| | |
|---|---|
| **TITLE** | **Data Aggregation Techniques for Internet of Things** |
| **COLLEGE** | **School of Communication and Information Engineering** |
| **AUTHOR** | **Sunny Sanyal** |
| **STUDENT NUMBER** | **L201620012** |
| **DEGREE CATEGORY** | **Master of Engineering** |
| **MAJOR** | **Information and Communication Engineering** |
| **SUPERVISOR** | **Prof. Dapeng Wu** |
| **SUBMISSION DATE** | **June 03, 2019** |

# Declaration of Originality

I declare that this thesis/dissertation is the result of an independent research I have made under the supervision of my supervisor. It does not contain any published or unpublished works or research results by other individuals or institutions apart from those that have been referenced in the form of references or notes. All individuals and institutions that have made contributions to my research have been acknowledged in the Acknowledgements. In addition, I understand that any false claim in respect of this work will result in disciplinary action in accordance with regulations of Chongqing University of Posts and Telecommunications.

Signature: 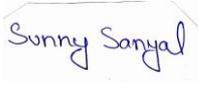          Date: 2019-04-09

# Copyright Permission Letter

I hereby authorize Chongqing University of Posts and Telecommunications (CQUPT) to use my thesis to be submitted to the CQUPT, namely I grant an irrevocable and perpetual license that includes nonexclusive world rights for the reproduction, distribution, and storage of my thesis in both print and electronic formats to CQUPT and to the relevant governmental agencies or institutes to reproduce my thesis. I also extend this authorization to CQUPT, for the purposes of reproducing and distributing single micro form, digital, or printed copies of the thesis or dissertation on demand for scholarly uses.

Signature of Student: 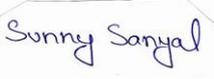          Date: 2019-06-03

Signature of Supervisor: 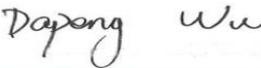          Date: 2019-06-03

# Abstract

With the widespread adaptation of Internet of things (IoT), we are already witnessing a deluge of IoT data analytic applications such as smart cities, air pollution monitoring and industrial supply chain management monitoring etc. IoT analytics can be defined as a process to control and optimize decision making in real time, by analyzing huge chunks of sensor data. The proper functioning of the data analytics demands high-quality data, low latency routing schemes, high energy efficiency and reasonable privacy. Given the requirements, a typical standalone data analytic framework fails to serve the purpose in practical scenarios. Moreover, the resource-constrained nature of IoT networks further aggravates the situation. To provide a sustainable solution; a data aggregation layer is introduced prior to the data analytics layer to increase the overall system efficiency. The data aggregation layer is responsible for efficient routing and data preprocessing.

The goal of this dissertation is to design efficient data aggregation frameworks for massive IoT networks in different scenarios to support the proper functioning of IoT analytics layer. This dissertation includes modern algorithmic frameworks such as non convex optimization, machine learning, stochastic matrix perturbation theory and federated filtering along with modern computing infrastructure such as fog computing and cloud computing. The development of such an ambitious design involves many open challenges, this proposal envisions three major open challenges for IoT data aggregation: first, severe resource constraints of IoT nodes due to limited power and computational ability, second, the highly uncertain (unreliable) raw IoT data is not fit for decisionmaking and third, network latency and privacy issue for critical applications.

This dissertation presents three independent novel approaches for distinct scenarios to solve one or more aforementioned open challenges. The first approach focuses on energy efficient routing; discusses a clustering protocol based on device to device communication for both stationary and mobile IoT nodes. The second approach focuses on processing uncertain raw IoT data; presents an IoT data aggregation scheme to improve the quality of raw IoT data. Finally, the third approach focuses on power loss due to communication overhead and privacy issues for medical IoT devices (IoMT); describes a prediction based data aggregation framework for massive IoMT devices.

# 摘要

随着物联网（IoT）的普及，出现了大量的物联网数据分析应用，如智慧城市、空气污染监测和工业供应链监管等。物联网分析可以定义为通过分析大量传感器数据实时控制和优化决策的过程。数据分析的正常运行需要高质量的数据，低延迟路由方案，高能效和合理的隐私。鉴于此，典型的独立数据分析框架无法在实际场景中实现这一目标。此外，物联网网络的资源受限性进一步加剧了这种情况。为了提供一个可持续的解决方案，在数据分析层之前引入一个数据聚合层，以提高系统的整体效率。数据聚合层负责高效的路由和数据预处理。

本文的目标是为不同场景下的大规模物联网网络设计高效的数据聚合框架，以支持物联网分析层的正常运行。本文包括非凸优化、机器学习、随机矩阵扰动理论、联合计算等现代算法框架，以及雾计算、云计算等现代计算基础设施。这种优秀设计的发展涉及许多开放性挑战，所提方案假设物联网数据聚合具有三大开放性挑战：第一、由于功率和计算能力有限，物联网节点受到严重的资源限制；第二、高度不确定（不可靠）的原始物联网数据不适合决策；第三、网络关键应用程序的工作延迟和隐私问题。

本文提出了三种独立的新方法，用于解决一个或多个上述开放性挑战的不同场景。第一种方法侧重于节能路由，讨论了基于固定和移动 IoT 节点的设备到设备通信的集群协议；第二种方法侧重于处理不确定的原始物联网数据，提出了一种物联网数据聚合方案，以提高原始物联网数据的质量；最后，第三种方法侧重于医疗物联网设备（IOMT）通信开销和隐私问题导致的功率损失，描述了一个基于预测的大型 IOMT设备数据聚合框架。

关键词：物联网，数据聚合，物联网分析，设备到设备通信，机器学习，优化，联合滤波

# Contents

# List of Figures

# List of Tables

# Abbreviations

| | |
|---|---|
| 3G | Third Generation |
| 4G | Fourth Generation |
| 5G | Fifth Generation |
| 3GPP | Third Generation Partnership Project |
| ARIMA | Autoregressive Integrated Moving Average Filter |
| AR | Auto Regressive |
| AM-DR | Adaptive Method of Data Reduction |
| AOA | Angle of Arrival |
| BS | Base Station |
| BI | Bussiness Intelligence |
| CQI | Channel Quality Indicator |
| CR | Communication Range |
| CM | Cluster Member |
| CH | Cluster Head |
| D2D | Device to Device Communication |
| DCM | D2D CQI Matrix |
| DS | Dominant Subspace |
| FFF | Federated Filtering Frameowrk |
| GPRS | General Packet Radio Service |
| GPS | Global Positioning System |
| IoT | Internet of Things |
| IoMT | Internet of Medical Things |
| LTE | Long Term Evolution |
| LTE-A | Long Term Evolution Advanced |
| LMS | Least Mean Square |
| MTC | Machine Type Communication |
| MPT | Matrix Perturbation Theory |
| PCA | Principal Component Analysis |
| RSSI | Received Signal Strength Indicator |
| RM | Relative Mobility |
| RPCA | Robust Principal Component Analysis |
| SVD | Single Value Decomposition |
| SGD | Stochastic Gradient Descent |
| TTI | Transmission Time Interval |
| UE | User Equipment |
| WiFi | Wireless Fidelity |
| WHO | World Health Organization |

# Chapter 1 Introduction

*"If you think that the internet has changed your life, think again. The Internet of Things is about to change it all over again"*- Brendan O'Brien, (Chief architect Aria Systems)

Over the past decades, the internet has undergone many incredible transitions to becoming an evolutionary technology. It all started in the early 90's with the internet of content; the web was static and is used to publish and share content. Eventually, the evolution resulted in the internet of services; where user-created content, XML, web services, productivity and commerce tools brought us better websites and services. Then the availability of affordable mobile broadband, smartphones and tablets brought us the internet of people; where people are connected to the world through social media, and mobile apps. Currently, we are standing on the intersection of an utterly disruptive technology enabled by upcoming 5G systems, Big Data analytics and Cloud/Edge computing; Internet of Things (IoT) [1].

IoT enables regular objects such as TV, lights and washing machine to become smarter by connecting them to the internet. This allows the devices to connect with each other and the environment efficiently. IoT is emerging as an unprecedented opportunity to transform the digital landscape for many players in the communications, information technology (IT) and consumer electronics industries. The World Economic Forum estimates; by 2025, 26-30 billion devices in homes and workplaces will be connected to the internet [2]. IDC expects a triple fold increase in global IoT market from 1.9 trillion US dollars in 2013 to 7.1 trillion US dollars in 2020 with 28.1 billion connected IoT devices [3]. Another similar forecast by Garter predicts the number of connected IoT devices will reach from 4.9 billion in 2015 to 25 billion in 2020 [4]. Apparently, the widespread adaption of IoT is increasing with passing time, and it will undoubtedly impact the future businesses and civilization.

The BI intelligence reports [5] that a typical urban land generates 4.1 TB of data per day per $Km^2$ starting 2016. The IoT applications reserve a significant portion in the total data generated from an urban land. Moreover, IoT applications are one of the key data producers in the urban scenario. Therefore the first logical step towards building any IoT based infrastructure should start with determining the properties of massive IoT data.

## 1.1    Taxonomy of Massive IoT data

Based on empirical observations [6] the IoT data is classified into two categories: data generation and data quality. A pictorial view of overall taxonomy is shown in Fig. 1.1.

### 1.1.1.    Data Generation

▪ Velocity: The IoT data is generated using various devices, and all the devices follow a different frequency of data generation. Some of the sensors such as accelerometers in sophisticated vehicles generate 10,000 readings every second. Other devices such as water monitoring sensors installed in canals, dams and water reservoirs scan the environment at a rate of 10 readings per hour. In cases where the rate of data generation is very high, the incoming data may overwhelm the system whereas a low scanning of IoT devices rate may lose important patterns in data

▪ Dynamics: The IoT data can be very dynamic in nature based on the source. A dynamic data source generates readings of different locations and times. As the source travels different environments and produces real-time data. More importantly, the source may not be moving through a well-connected path, and as a result of intermittent connectivity, IoT data suffers many irregularities.

▪ Heterogeneity: The IoT paradigm has the potential of connecting everything to the internet. This 'everything' includes jewellery, shoes, vehicles, classrooms etc. Therefore, the data from diverse IoT devices follow different formats, modality and sizes.

### 1.1.2  Data Quantity

▪ Uncertainty: The uncertainty in IoT data refers to the fact that raw (unprocessed) IoT sensor data is highly untrustworthy and definitely not fit to be used by any decision making systems. The uncertainty in the raw IoT data is due to the lack of precision in the IoT device that causes missing values, outliers and wrong values. Noise is also a major uncertainty.

- **Redundancy**: Redundancy in IoT data refers to the fact that there are multiple copies of the same data in the IoT sensor dataset. Redundancy costs a lot of processing power and storage that diminishes the overall system efficiency. Redundancy may occur in various scenarios such as multiple IoT devices are deployed in the neighborhood to sense a particular phenomena.

- **Ambiguity**: The particular IoT dataset can be perceived in many different ways depending upon the scenario and the consumer requirements. Proper interpretation of data is an essential condition for increasing the reusability of data. In some cases where there are multiple possible usages of a particular dataset, ambiguity becomes a significant challenge in proper interpretation of data.

- **Inconsistency**: Inconsistency is prevalent in massive IoT sensor data. Inconsistency is caused by poorly calibrated or damaged IoT device. Due to environmental hazards, less skilled workforce and various other unavoidable circumstances the IoT device generates random readings that are not an accurate representation of the phenomenon it is responsible for sensing.

## 1.2  IoT Analytics

The deluge of data or simply massive data generated by IoT applications must be analyzed to fetch information for decision making. IoT analytics can be defined as a process to control and optimize decision making in real time, by analyzing huge chunks of big data. The IoT analytics is required to analyze every segment of the maasive IoT data to obtain crucial patterns in the data stream. It can help industries in preventive maintainance for equipments and other infrastructures, it also helps them in adopting new business models, streamline operational processes, generate novel products and services for users (refer Fig. 1.2). It is worth mentioning that a clear distinction between IoT analytics and big data analytics; the IoT analytics does not deal with all the characteristics of big data (5 V characteristics; volume, velocity, veracity, value and variation).

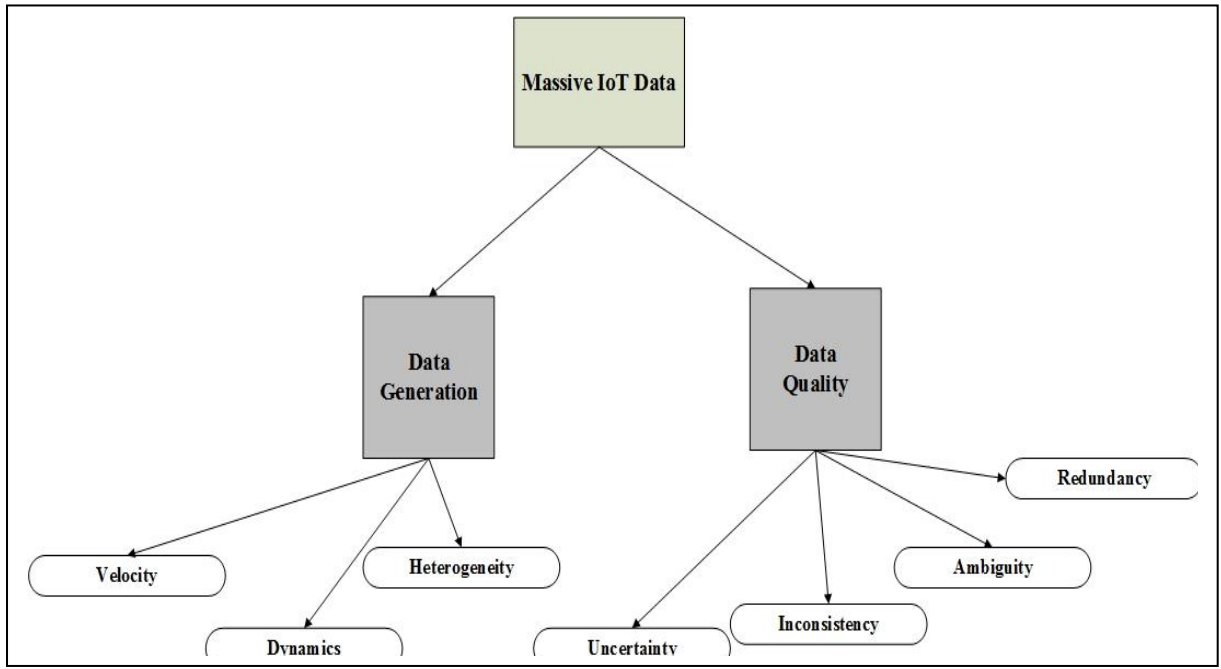

Fig. 1. 1 Massive IoT data Taxonomy

In particular, three sectors are more likely to be effected by the imminent IoT analytics paradigm.

- Smart lifestyle: based on the predictions made by Gartner by 2022 each family will possess more than 500 smart devices. Another similar prediction [7] states that by 2020, the connected kitchen will contribute to at least 15% savings in the food and beverage industry using big data analytics.

- Smart transport: Garter predicts [8] by 2020 a quarter million connected vehicles will facilitate new vehicular and autonomous driving competency. The connected vehicles will generate 30 TB of data each day while communicating with the environment and other vehicles. This new paradigm will generate a business potential worth 14 billion US dollars worldwide.

- Smart cities: San Jose, USA, aims to improve quality of life through air quality monitoring, traffic flow and more [9]. In Pisa, Italy, intelligent guidance helps drivers to find free parking space [10].

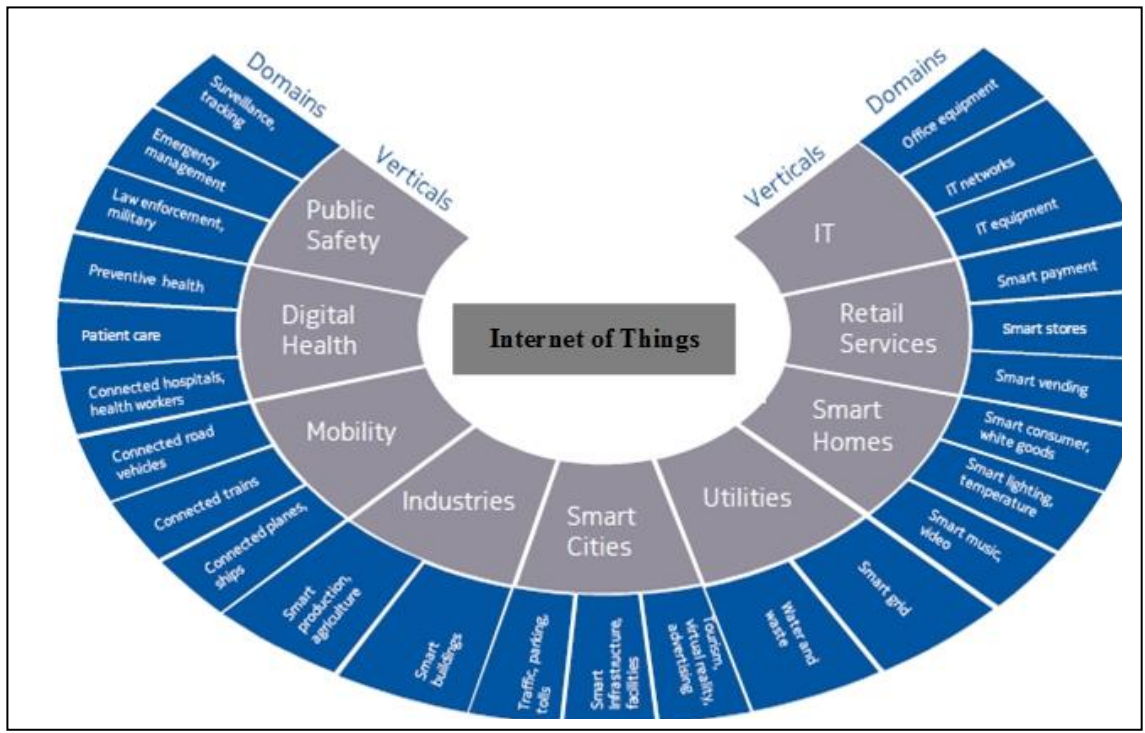

Fig. 1. 2 Map of IoT domains and verticals

As discussed IoT analytics have proven to bring value towards the society, and currently it is receiving considerable attention from both academia and industry. The growing interest in IoT analytics need its stakeholders to clearly understand the approaches involved in analytics, building blocks, technical requirements and open challenges.

## 1.2.1  IoT Analytics Architecture

A state of art IoT analytics framework is aimed at extracting information out of massive IoT data and performing decision making tasks based on the fetched information. The entire IoT analytics network can be classified into three major parts, as depicted in Fig. 1.3. The sensor networks are sensing the environment and generating readings. These networks are also responsible for sharing the data with routing device or gateway optimally regarding the power consumption, communication overhead and shortest possible path. Furthermore, the routing network can be a gateway device or an IoT device acting as a cluster head for a cluster of IoT nodes delivering the aggregated data to the analytics server.

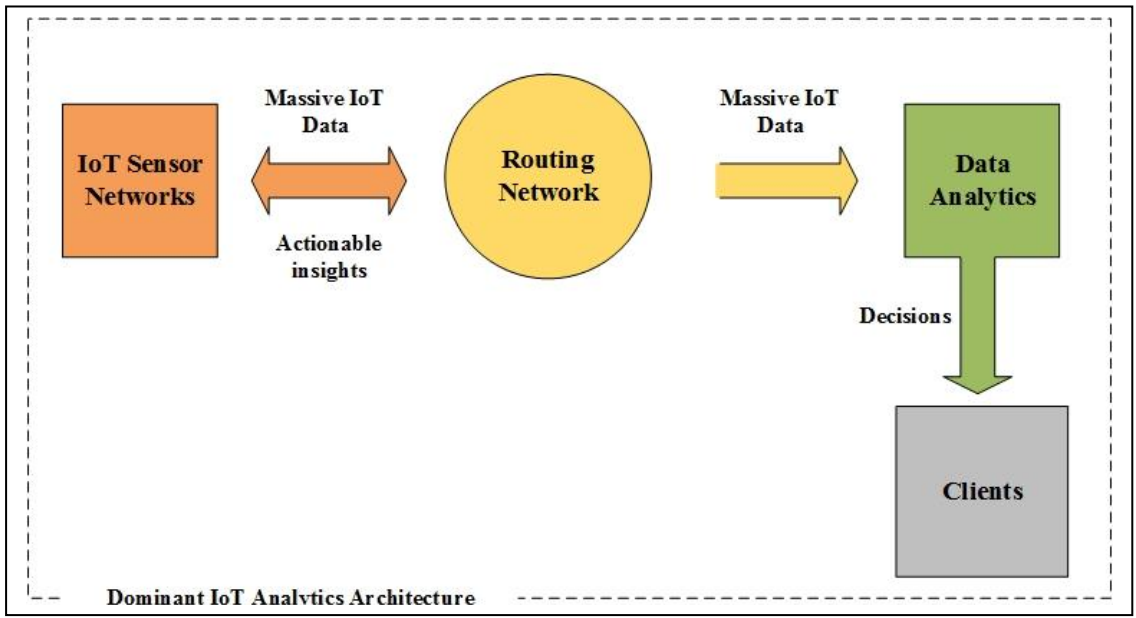

Fig. 1. 3 Dominant IoT Analytics Architecture

Typically, the analysis and decision making tasks of massive IoT analytics platform are performed inside the cloud servers in a centralized fashion. A centralized IoT analytic system (dominant) demands all the data at once to make inferences.

## 1.2.2  Taxonomy of IoT Analytics

Extracting business value out of the raw IoT sensor data is anything but trivial. In order to leverage the right data at the right moment to cater to the client requirements it is crucial to categorize IoT analytics based on application scenarios and respective use cases. A comprehensive taxonomy [11] of IoT analytics is presented pictorially in Fig. 1.4. IoT analytics can be broadly classified into two categories: historical analysis and proactive analysis. The historical analytics provides a comprehensive visual interpretation of the data. It is generally based on traditional data mining techniques. The historical data analytics can be further classified into descriptive analytics and diagnostic analytics. The descriptive analytics generates statistics and visual interpretation of data and it is generally used in businesses to evaluate their products and user base. The diagnostic analytics performs anomaly detection to check malfunction in equipments and provides an alarm in the case on any occurrence of an anomaly.

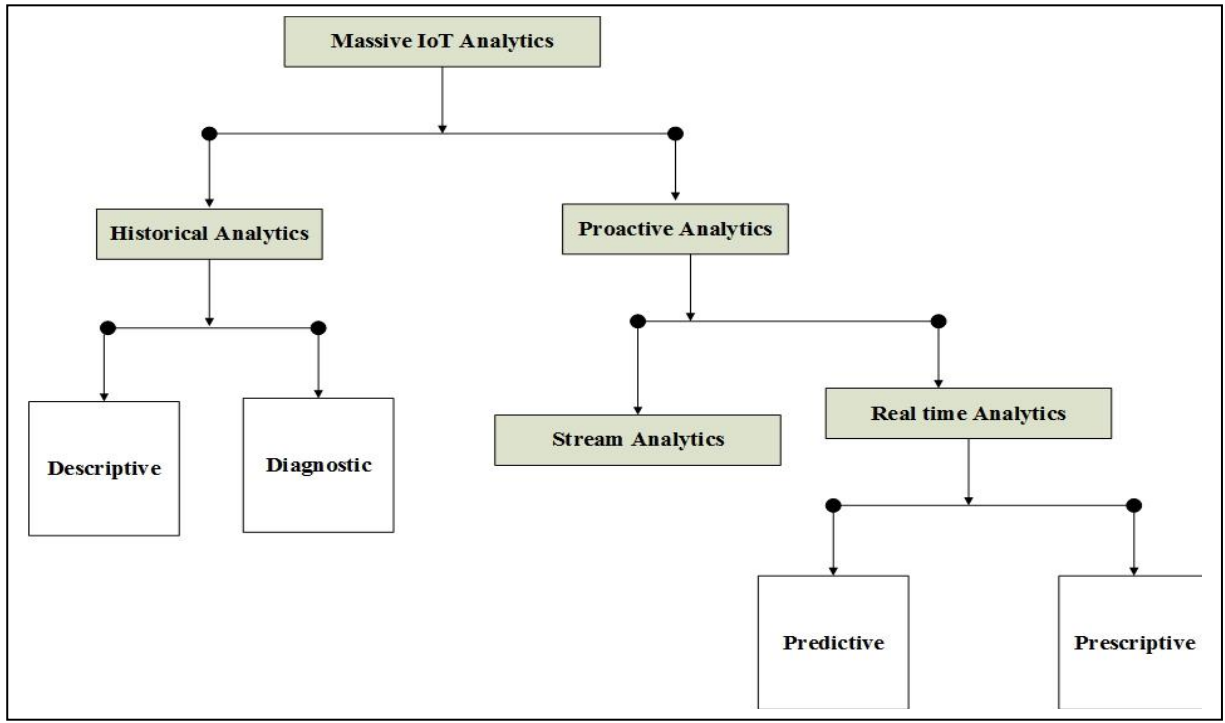

Fig. 1. 4 Diffirent Categories of IoT Analytics based on use cases

On the other hand, proactive analysis is slowly but steadily emerging as a new trend to generate and facilitate actionable insights from the massive IoT dataset. This category is further classified into stream analytics and real-time analytics. In the case of stream analytics, the incoming time series data is analyzed based on batches or streams; the idea is to find a vital pattern inside the batch or stream of the dataset. The real-time analytics is required to generate optimal or suboptimal output by analyzing a part of data (not the full dataset) within a minimal time budget. The real-time analytics is getting popular due to its wide range of applications in civil society and industry.

## 1.3   Open Challenges

The proactive decision making is swiftly becoming the mainstream since it can cater the dynamic applications for industries and present day society. However, despite all the hype about data analytics real time IoT analytics is still in its infancy. The real-time IoT analytics infrastructure has several challenges, and proper functioning of the system demands high-quality data for decision making, efficient low latency routing schemes, overall high energy efficiency, optimal communication overhead and reasonable privacy. Moreover, the resource-constrained nature of IoT networks further aggravates the situation. The dominant architecture of data analytics as shown in Fig. 1.3 is not competent to support proactive analytic applications that require real-time analysis. An evident problem with the dominant data analytics paradigm is that the analytics layer is standalone. Although this paradigm has worked well for historical data analysis, the modern-day proactive analysis demands online decision making within a limited time budget. To discuss and evaluate the severity of the open challenges involved in each layer of the dominant IoT analytics architecture (Fig. 1.3); this dissertation classifies overall setbacks into three categories: open challenges related to sensor networks, open challenges related to the routing network and open challenges related to data analytics.

## 1.3.1  Open Challenges related to Data Analytics Layer

The Data veracity problem is one of the primary concerns for any data analytics system. The unprocessed (raw) IoT sensor data is highly uncertain. The uncertainty is caused due to the presence of outliers, missing values, redundancy, inaccuracy and biased readings. The outliers are readings of a dataset that show a complete aberration from an established pattern of a data batch. The missing values adversely affect the machine learning algorithm used by analytical platforms for decision making and in most of the cases it results in random outputs. Redundancy poses threat to the system level efficiency. Typically a system overwhelmed with redundant data losses great portion of its computational resources to process duplicate data at multiple occasions. Hence, redundancy reduces overall system efficiency. Finally, inaccuracy and bias are very hard to detect in a massive IoT dataset as it is context dependent and to identify the anomaly the system or an individual must have considerable domain expertise. Overall the data veracity problem poses a significant threat to all applications involving IoT analytics platforms.

Traditionally the IoT analytics/decision making happens inside the remote cloud servers or cloud data centers. The cloud servers are embedded with fully centralized decision making algorithms that demand access to the entire dataset in one go. Therefore such systems require continuous pushing of all the massive data generated by IoT networks to the cloud servers; which causes several critical challenges regarding scalability, network latency, network energy efficiency and sensitive privacy.

The IoT device individually generates small packets of data very frequently; considering the size of overall networks, the small data chunks collectively become massive in volume. Moreover, the massive IoT data volume is complemented with a very high rate of generation that makes it streaming time series sensor data. Hence scalability is a crucial requirement for real-time IoT analytics application. In practical scenarios, the streaming massive IoT data can choke/overwhelm any system. The traditional cloud computing paradigm has proven its worth in historical data analytics for more than a decade; however, it is not suitable to support real-time IoT analytics application such as industrial IoT analytics and healthcare analytics.

The network latency issues arise from a number of different scenarios in the analytics network architecture. The network delays may occur due to frequent long-distance transmissions of data, interlayer communication gap in a tier analytics architecture, intermittent communication and inefficient communication subroutines for decision-making task etc. The foremost issue identified for network latency especially in suburban and rural areas is the long-distance transmission of data to the remote cloud servers in big cities. For example, the massive IoT network of 100,000 sensors deployed in a 14, 00 Km waterway from Beijing to Tianjin is continuously monitoring the water quality. The sensor deployed near the Tianjin needs to transmit data to Beijing as the cloud servers are located in Beijing; hence the system incurs significant delays [12]. In a nutshell, network delays can be detrimental to proactive IoT analytics.

The IoT analytics paradigm is emerging as a critical driver in smart healthcare solutions. In scenarios related healthcare the data is highly sensitive in nature and anonymity of the data is the primary concern of the clients. The dominant centralized cloud computing paradigm aggregates all the data in public or private servers vulnerable to cyber attacks; therefore, it is highly unsuitable for such critical applications since it may lead to exposure of privacy.

### 1.3.2  Open Chanllenges related to IoT Sesnor Networks

Adding semiconductors and radios to things that previously had none drastically reduces the life span of the product [13]. Typically IoT devices are small in size, and the battery is even smaller; due to this design constraint, many IoT devices are disposed once it is out of charge. Therefore IoT devices have severe resource constraints regarding power, computation and storage abilities. Increasing the battery life of IoT devices is a crucial problem, and it is yet to be solved. Therefore while performing decision making based on machine learning techniques IoT devices requires an entirely different approach although; the traditional machine learning algorithms require higher resources that are not available to these small IoT devices. The problem of performing IoT analytics in a resource-constrained scenario brings significant challenges to both the embedded software developers and network designers.

### 1.3.3  Open Chanllenges related to Routing Network

Cooperative communications are going to be an important asset to the imminent 5G communications. Cooperative communications also known as device to device communication (D2D). Ineffective D2D communications leads to many issues as high power loss, high communication overhead and high packet loss. Moreover, in dynamic scenarios where IoT devices are non-stationary, the routing network suffers greater power loss and packet loss that further aggravates the IoT data routing. Therefore IoT routing network requires better schemes that may provide high performance for stationary and dynamic IoT scenarios.

## 1.4  Potential Solution

As discussed before the dominant IoT analytics architecture burdens the data analytics layer with several responsibilities. These responsibilities include tasks required for decision making and other supporting tasks. However, in the process of balancing the tradeoff between overall system efficiency and decision making, the data analytics layer compromises the performance of decision making. To provide a sustainable solution; a data aggregation layer is introduced before the data analytics layer that increases the overall system efficiency. Broadly the data aggregation layer is responsible for efficient routing and data preprocessing. It facilitates the data analytics layer to entirely focus on decision making while it performs the tasks required to support IoT analytics such as optimal collection of data schemes, ensuring the quality of data, maintaining energy efficient low latency communication and securing data privacy. Apparently, the aggregation layer downloads the supporting tasks from data analytics layer. The proposed IoT analytics architecture (refer Fig. 1.5) facilitates data analytics layer to perform effective decision making by supplying all the necessary resources to sustain an overall near optimal system-level performance. The proposed architecture involves an amalgamation of modern technologies such as fog computing, cloud computing and machines learning.

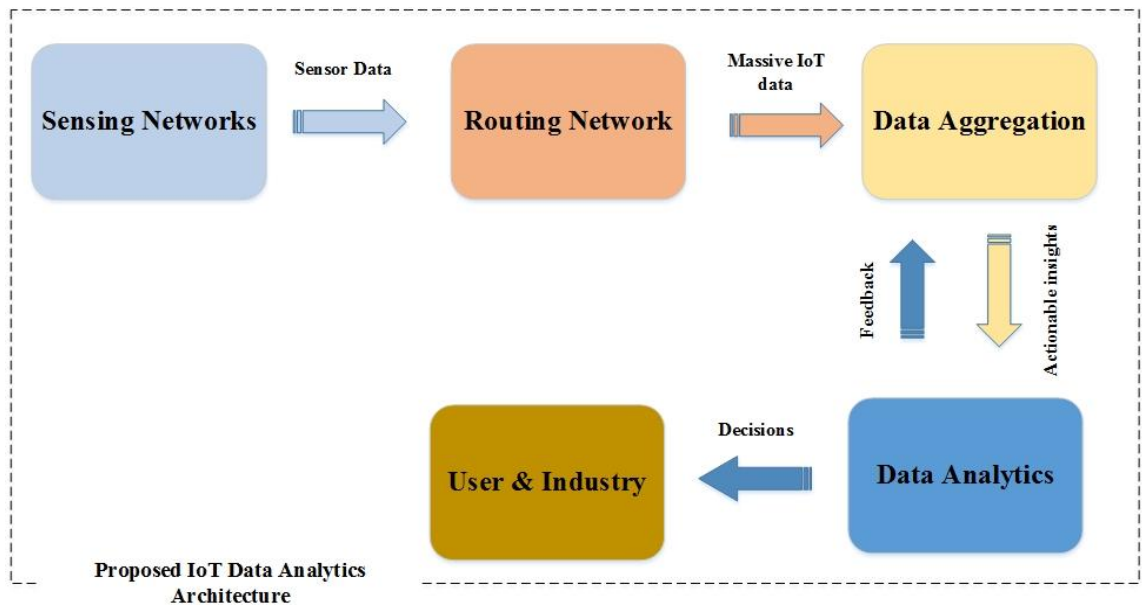

Fig. 1. 5 Proposed IoT Analytics Architecture

## 1.4.1   Objective

The core objective of this dissertation is to design efficient data aggregation algorithms and frameworks for massive IoT networks in diverse scenarios to support the proper functioning of the entire IoT analytics sytem. To achieve this goal, the dissertation investigates data-driven approaches for massive IoT data aggregation that rely on methods based on nonconvex optimization, federated learning framework and machine learning.

The data aggregation is highly dependent on the operational environment; for example, the dynamics of data aggregation inside underground mines has a distinct characteristic than data aggregation on the ground surface. To achieve the primary goal while solving the related open challenges involved this dissertation investigates different scenarios as follows:

- ▪ The dissertation investigates a complex data delivery network scenario having both stationary and non-stationary IoT devices. The goal of this part of the investigation is to present and validate a unified mobility based aggregation scheme based on D2D communication for stationary and non-stationary IoT devices. The mentioned scheme is also very energy efficient as the IoT devices are assumed to be severely resource constrained.

- ▪ Another ubiquitous scenario receiving much attention is an analytic system receiving highly corrupted massive IoT sensor data as input. The objective for this part of the investigation is to solve the data veracity problem for raw IoT and to improve the quality of extensive raw IoT sensor data.

- ▪ The dissertation investigates a very popular and crucial application sector of IoT analytics, i.e. smart healthcare. The central goal of this investigation is to increase device battery life by reducing the communication overhead and maintaining reasonable privacy for sensitive medical data.

## 1.4.2 Overview

This dissertation presents three distinct senarios for massive IoT data aggregation based on the proposed data analytics architecture to solve one or more open problems discussed in section 1.3. The approaches are briefly summarized:

- ▪ **Mobility based cooperative data aggregation scheme:** Routing is an integral part of data aggregation as it includes both the IoT sensor networks layer and the aggregation layer. A well-routed energy efficient delivery scheme not only prolongs the life of the IoT networks but also reduces latency and packet drops. This approach, focuses on the wireless domain, where energy remains a key constraint. To solve this problem, this dissertation proposes a data routing framework, whose application can be extended to IEEE 802.15.4 protocol, i.e. low power devices (IoT devices). The idea is to amalgamate the resource-constrained IoT delivery network model with mobility based energy efficient clustering utilizing D2D communications. The literature of IoT data routing using D2D communication lacks a unified approach that is applicable for both static and dynamic IoT devices; this part of the dissertation intends to fill this gap through this research work. The simulation results depict that our heuristically developed scheme is highly energy efficient.

- **Improving the quality of massive IoT sensor data:** Data analytics stands on the shoulders of sensor data aggregation that includes data pre-processing and routing. This portion of the dissertation envisions two open problems of data aggregation; first, raw IoT sensor data is highly uncertain; second, the traditional algorithms are not fit for processing highly uncertain sensor data. This is formally known as data veracity problem. This part of the dissertation proposes a data aggregation scheme for highly uncertain raw IoT sensor data collected using D2D communication. The mentioned scheme initially reconstructs the subspace using sample data, and then it iteratively tracks down the low-rank approximation of the dominant subspace in the presence of high uncertainties at the fog server. Later the robust dominant subspace is used to estimate a more reliable true sensor data matrix from the highly uncertain raw IoT sensor data traffic matrix. Moreover, the proposed scheme achieves the tasks while processing the raw IoT sensor data without any prior information, i.e. in a fully unsupervised fashion. The existing literature based on sampling, approximation and data reduction either causes random data reduction or destruction of global characteristics of the raw data. However, unlike the existing solutions, the proposed method removes uncertainties while preserving the global characteristics of the raw data. Performance evaluations conducted using both the real world sensor data and synthetic data injected with noise, outliers and missing values. Experimental results show that the proposed approach can estimate a reliable true sensor data matrix in the presence of high uncertainties and the energy efficient device to device communication-based data delivery mechanism can accommodate a large number of IoT devices.

- **Federated Filtering for IoT medical data aggregation:** Based on the dominant paradigm, all the wearable IoT devices used in the healthcare sector also known as the internet of medical things (IoMT) are resource constrained in power, storage and computational capabilities. The IoMT devices are continuously pushing their readings to the remote cloud servers for real-time data analytics that causes faster drainage of the device battery. Moreover, other demerits of continuous centralizing of data include exposed privacy and high latency. This particular portion of the thesis presents a novel Federated Filtering Framework for IoMT devices which is based on the prediction of data at the central fog server using shared models provided by the local IoMT devices. The fog server performs model averaging to predict the aggregated data matrix and also computes filter parameters for local IoMT devices. Two significant theoretical contributions of this approach are the global tolerable perturbation error ($Tol_F$) and the local filtering parameter ($\delta$); where the former controls the decision-making accuracy due to eigenvalue perturbation and the later balances the tradeoff between the communication overhead and perturbation error of the aggregated data matrix (predicted matrix) at the fog server. Experimental evaluation based on real healthcare data demonstrates that the proposed scheme saves upto 95 % of the communication cost while maintaining reasonable data privacy and low latency.

### 1.4.3  Major Contributions

This dissertation focuses on algorithms and system development for IoT massive data aggregation to solve one or more open challenges. One size fits all strategy miserably fails to support data aggregation for IoT networks as this process highly dependent on the operational environment and the dynamics of the system. The overall contribution of this dissertation is the integration of IoT data aggregation layer with compelling machine learning techniques. Moreover, this dissertation is also credited with the frameworks for crucial applications that integrates the fog computing and modern cloud computing paradigm with IoT networks. The individual contributions of each chapters are explicitly stated in the chapter overview.

## 1.4.4  Thesis Organization

The remainder of the thesis is organized in the following fashion. Chapter 2 discusses mobility based cooperative data aggregation approach. This chapter is directed towards presenting a multihop cooperative communication-based data aggregation scheme for stationary and non-stationary IoT nodes. Initially, it discusses the cluster formation, and delivery scheme and next simulates a dynamic IoT network scenario. This dissertation then discusses a data aggregation scheme for highly uncertain massive IoT sensor data and describes the formulated optimization problem in Chapter 3. The chapter also presents a comprehensive solution to the optimization problem and also an algorithm that implements the theoretical solution into practice. In chapter 4 the dissertation presents a data aggregation framework for the internet of medical things (IoMT). This chapter has two main components; first, the federated filtering framework and second, theoretical upper bounds on tradeoff for local filtering and global perturbation of the centralized data. Finally, the dissertation summarizes lessons learned, outline future work and conclude in Chapter 5.

# Chapter 2 Mobility Based Cooperative Data Aggregation Scheme

## 2.1    Chapter Overview

The advent of the internet of thing brings a lot of new opportunities and research avenues for the world. Today IoT has become a favourite research topic among most all the top tech companies on the Forbes list of the new startups. This era will be remembered as a stepping stone to a more connected and conscious world where IoT has implementations from space ships to body cells. However to leverage benefits out of the IoT paradigm it's crucial to develop efficient mathods to route the data generated by the myriad of IoT applications in the physical layer. The coexistence of diverse communication technologies (3G/4G/5G/Wi-Fi/GPRS) at the same time, spare no other option but to find a cooperative technology that can be used to route data with the help of peers in the proximity. 3rd Generation Partnership Project has defined the aforementioned technology as machine device communication (D2D) instead of MTC. The device to device communication between user equipment (UEs) in proximity enhances the spectral and power efficiency.

This chapter discusses a data aggregation scheme for both stationary and nonstationary IoT nodes with minimal resources regarding power and computational capabilities. Based on the proposed data analytics architecture discussed in chapter 1 the IoT sensor networks layer is controlled and optimized by the data aggregation layer to aggregate the massive IoT data effectively. The chapter identifies power budget, device dynamicity and computational complexity among UEs (IoT devices) as the major open challenges. The device densities of the UEs in specific locations is utilized in the proposed algorithm for D2D communication in mutual proximity, the kernel of the scheme is to cooperatively upload the content of UEs to the remote base station (BS) by forming a multi-hop D2D framework. Initially, the devices form cluster having a cluster head at the helm. The cluster heads form a multi-hop path among themselves to upload the data to the BS. Moreover, the data is uploaded by that cluster head which is closest to the BS. In this chapter, our work is focused on developing energy efficient and computationally less complex algorithm for cooperative IoT data aggregation.

Major contributions are listed below:

- Due to the limited power budget the use of GPS is not a prudent option. A novel alternative is proposed to compute the velocity of non-stationary IoT device. As discussed in section 2.3.2.
- The algorithm I (section 2.4) is a lightweight local subroutine for cluster formation, this heuristic approach is highly energy efficient.
- The algorithm II (section 2.5) the proposal presents an energy efficient data aggregation is based on multihop inter-cluster D2D communication.
- The proposed approach accommodates both stationary and non-stationary IoT devices.

## 2.2    Background and Related Work

The Device to device communication was first introduced in 3GPP [15] in release 11 and later to bolster the framework further technical details were published in release 12. The Proximity communication enabled by D2D has gained huge momentum as a way to overcome the demerits of the conventional cellular system. The major merits of D2D communications are: (a) Low power consumption, due to interaction of devices in mutual proximity requires less power, (b) transmissions with high data rate due to mutual cooperation among devices, (c) reliable communication, (d) lower overhead to BS and (e) heterogeneous connectivity of devices, since devices using different technologies such as Wi-Fi, LTE-A can be accommodated.

D2D can be categorized in two different ways [16] one is implemented using a licensed cellular spectrum (inband D2D) and the other uses unlicensed spectrum (outband D2D). The inband communication is classified into two categories: underlay D2D where, cellular and D2D communication share the same resources simultaneously whereas, in overlay D2D, there are dedicated resources for D2D to operate. The outband D2D aims to eliminate the interference problem, but since BS is not involved in the scenario, it needs to rely on interfaces like WiFi, Bluetooth and infrared etc. D2D technology has been read and researched for quite some time, and many published research articles successfully depict the general idea and modern implementations of the technology.

Due to the vast number of IoT devices, D2D is the favourite candidate to be used for cooperative data aggregation. A. Orsino et al. in their paper [16] have presented a fascinating scenario of IoT data aggregation for smart cities; this paper has addressed a wide range of problems with a detailed solution. However, some assumptions like stationary UEs, round-robin resource allocation etc, limit the scope of the solution. G. Rigazzi et al. have also proposed an excellent alternative resource allocation scheme in that is both practical and easily implementable. Militano et al. in the paper [17] have presented a cooperative coalition formation based on social trust and proximity.

Data aggregation layer can also be used to manage the resource block (RBs) and power efficiently, the literature of this technology is vast. Moreover, a good taxonomy of data aggregation can be found in [18] in which Schubert et al. have proposed a novel data aggregation technique for IoT devices, they used data aggregation as a tool to reduce the power consumption. In general, we observe that the wireless research fraternity is trying to deal with the IoT data transmission the same way as they dealt with wireless sensor networks data transmission. IoT data transmission has certain differences with wireless sensor network (WSN) data transmission: (a) the content generated by IoT devices are relatively smaller than general WSN data, (b) generally the channel is expected to be good for data transmission in WSN but in IoT it doesn't make much difference, (c) in IoT data transmission the system latency should be minimum unlike WSN where generally there is no such obligation. Finally, the aforementioned related works are focused only on reducing power consumption of IoT data transmission; however, the solution proposed in this chapter considers a practical tradeoff between computational complexity and power consumption in IoT data transmission using D2D communication.

## 2.3    System Model

The proposed system is formulated using Long Term Evolution Advanced (LTE-A) network, which in case provides dedicated resources for D2D communication and the system follows inband overlay D2D communication protocol. To maintain the simplicity this chapter demonstrates our proposal using a single cell scenario as shown in Fig. 2.1.

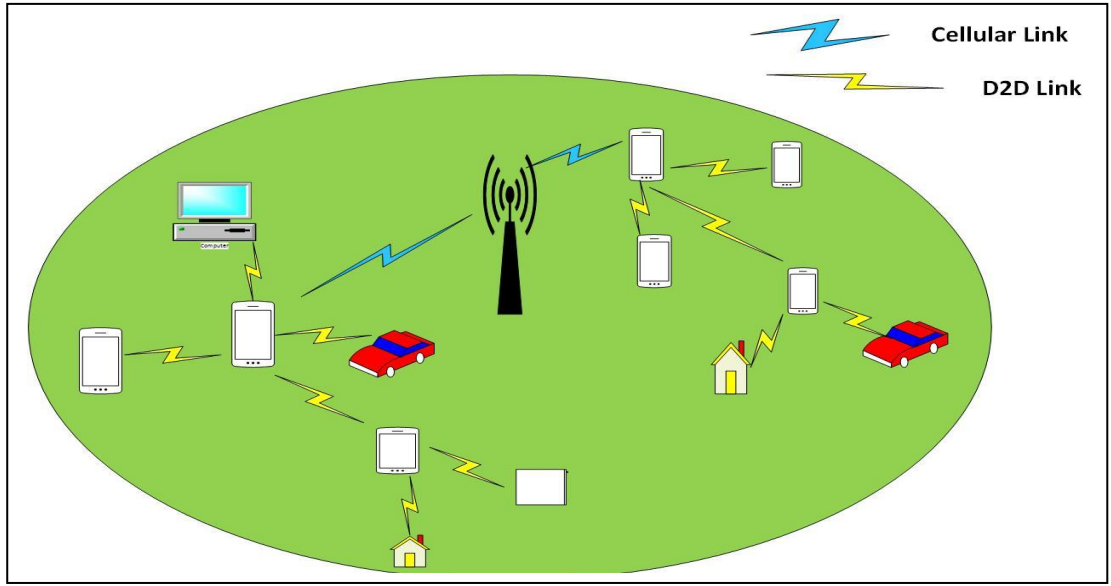

Fig. 2. 1 Proposed IoT Analytics Architecture

Following the standard procedure of data uploading in the cellular network, primarily the BS collects the cell-mode CQI (channel quality indicator) feedbacks from all IoT devices (nodes) willing to upload some data. Secondarily the BS collects the D2D mode CQI values and forms a D2D CQI matrix (DCM) [16]. Suppose each node $n_i = \{n_{1,\ldots\ldots},n_{|n_i|}\} \in N$ is a part of cluster $s_i = \{s_{1,\ldots\ldots},s_{|s_j|}\} \in S$ in a cell governed by a BS. Both the stationary and non-stationary IoT devices (nodes) broadcast beacon messages to find a 1-hop neighbour. Here the participants use a range based method to evaluate the angle of arrival (AOA) and received signal strength indicator (RSSI) [19] to locate the distance of an unknown neighbouring node and angle w.r.t the reference node (node which needs a relay node to transfer data). The co-relative mobility (discuused in section 2.3.3) between nodes is calculated and stored in each node as a matrix. Next, all the nodes select a 1-hop neighbour node based on co-relative mobility among the neighbouring nodes to relay their data. This process when implemented by all nodes in a cell, the nodes group among themselves to from dynamic or static clusters.

This chapter assumes that each device is a smart device with either accelerometer or GPS mounted inside it. Based on the underline principle of inband overlay device to device communication BS provides dedicated resources. Moreover using the DCM (D2D CQI matrix) mentioned previously, BS monitors and controls entire D2D scenario. A link that is not feasible as per the DCM is not connected.

As the algorithm runs the BS allocates node ids to each node and makes rings of nodes based on the distance from it. Then, each node in the cell broadcasts beacons to locate a relay node in its proximity. A relay node is an IoT device that helps a source node to upload their data cooperatively to the BS using D2D links. Here the devices use a range based method to evaluate the angle of arrival (AOA) and received signal strength indicator (RSSI) to locate the distance of an unknown neighbouring node and angle w.r.t the reference node (node which needs a relay node to transfer data). Each node computes a dynamic N x N matrix that is similar to DCM matrix however instead of CQI value it contains co-relative mobility of reference node w.r.t all other nodes. This matrix further plays a crucial role in searching a deserving neighbouring node to relay data. This chapter also assumes that all the UEs are smart devices and can send or receive data autonomously. Prior moving to the main proposal we need to define certain keys concepts that is used throughout the proposal.

## 2.3.1   Energy Efficiency

This chapter focuses all the efforts to optimize the energy efficiency with the computational complexity, which in case is a trade-off between these two parameters in this proposal. Moreover, the amount of energy to be used for the proposal is based on the number of packets to be uploaded by each node. Since energy efficiency ($\eta$) is researched for a long time the proposal uses standard equation [16] from the literature. The energy efficiency ($\eta$) of the system which has N number of users can be computed as:

$$\eta = \sum_{N} \frac{d_N}{E_N \cdot r_N \cdot TTI} \qquad (2.1)$$

Where $d_N$ is the total volume of data (in bits) to be uploaded, $E_N$ is the average power consumed to deliver a single packet of data, $r_N$ is total number of data packets to be uploaded by all the nodes and TTI is transmission time interval which is fixed for all packets.

### 2.3.2  Accelerometer

Accelerometer is a device that measures the acceleration and is omnipresent in the IoT devices globally. A 3 axis accelerometer in IoT device provides the X, Y, Z coordinate values, which is used to measure the position and acceleration of the device. The rotation, direction and position are measured using gyroscope sensors. The proposal is only interested in linear acceleration of the device. The accelerometer reading is filtered to generate useful results. The function that calculates velocity is shown below.

$$v(t) = v(0) + \sum a \times \delta t \qquad (2.2)$$

Where v (t) is the instantaneous velocity at time t, v (0) is the initial velocity, a is the instantaneous acceleration at time t and $\delta t$ is the time taken to accelerate.

Here the proposal is interested in the horizontal component of velocity of nodes. The horizontal component of velocity of node $n_x$.

$$V_x(t) \ \text{c}\ \alpha s \qquad (2.3)$$

Where $\theta$ is the angle of velocity w.r.t other node. The proposal is focused on reduction of power consumption that increases the device battery life. Therefore, the chapter advocates the use of accelerometers as speedometers to calculate the velocity of resource constrained IoT nodes. This concept saves higher energy and computations otherwise used by GPS to get the position coordinates and then compute the velocity based on movement. The accelerometer readings may not be accurate but since the proposal deals with the relative velocity, it is suitable for the purpose.

### 2.3.3  Co-relative Mobility

Relative mobility [20] has a vast literature and research fraternity have used the term relative mobility in fairly different scenarios. Moreover in this proposed scheme the term co-relative mobility is used as a function i.e. cross correlation of relative mobility between two nodes $n_x$ and $n_y$ where $x, y \in i$ and $n_x, n_y \in N$ at time t. As already mentioned to calculate the distance between two nodes, the proposal uses the RSSI and AOA techniques. Distance between two nodes $n_x$ and $n_y$ where $x, y \in i$ and $n_x, n_y \in N$ at time t can be defined as the $D_{XY}(t)$. The normalized distance $\bar{D}_{XY}(t)$ can be defined as:

$$\bar{D}_{XY}(t) = {D_{XY}}\big/{CR} \qquad (2.4)$$

Where CR is the maximum communication range of the beacon. The horizontal velocity of nodes can be defined as:

$$V_x(t)\cos\theta \qquad (2.5)$$

$$V_y(t)\cos\theta \qquad (2.6)$$

Here $\theta$ is the velocity vector angle, moreover the orientation of the movement w.r.t the reference node can be measured by $\theta$.

$$\bar{V}_x(t) = |V_x(t)\cos\theta| \qquad (2.7)$$

Where $\bar{V}_x(t)$ is only the magnitude of the horizontal component of velocity of node $n_x$.

$$RM_X(t) = \alpha\bar{D}_{XY}(t) + \beta\bar{V}_X(t) \qquad (2.8)$$

Here, $RM_X(t)$ is the relative mobility of node $n_x$. The values of $\alpha$ and $\beta$ are subjected to change based on traffic scenarios. Using eqs. (2.4), (2.7) and (2.8) the co-relative mobility between node X and node Y can be defined as the cross correlation between the relative mobility of two nodes as shown below:

$$cr_{XY}(t) = Correlation(RM_X(t), RM_Y(t)) \qquad (2.9)$$

Here, $cr_{XY}(t)$ is the co-relative mobility of two nodes $n_x$ and $n_y$ where $x, y \in i$ and $n_x, n_y \in N$ at time t and n is the total number of nodes.

## 2.4    Proposed Cluster Formation Scheme

The formation of cluster can be summarized:

- ▪ The BS randomly allocates a unique node id to all the nodes.

- • The BS (located at the centre of the cell) groups the IoT nodes in several circular rings based on the radial distance from the center. The BS also allocates ring IDs to every IoT node as per the position.

- • The participating nodes (reference nodes) broadcast advertisement messages within their radial range of transmission. Simultaneously the neighbouring nodes receiving the messages within the radial range compute the co-relative mobility with the advertiser node based on Eq. 2.9.

- • The neighbouring node computes the competency of the reference nodes (advertiser node) to join a cluster based on the criteria of co-relative mobility. The neighbouring node stores this competency value of each advertiser nodes in an array known as competency array. This array contains competency values of the nodes interested in relaying data.

- • The neighbouring node (current node) joins a cluster that has the highest co-relative mobility among the other advertiser nodes, which is also the most competent relay node for the advertiser node. An important constraint regarding the grouping of IoT nodes is that a node can only join a cluster where the cluster head has the same ring ID.

- • When the process mentioned above is implemented over a large group of nodes, they form several clusters of variable sizes.

The process mentioned above can be easily understood using the flowchart shown in Fig. 2.2. The following section depicts our idea of data uploading to BS based on the inter-cluster multi-hop data transmission.

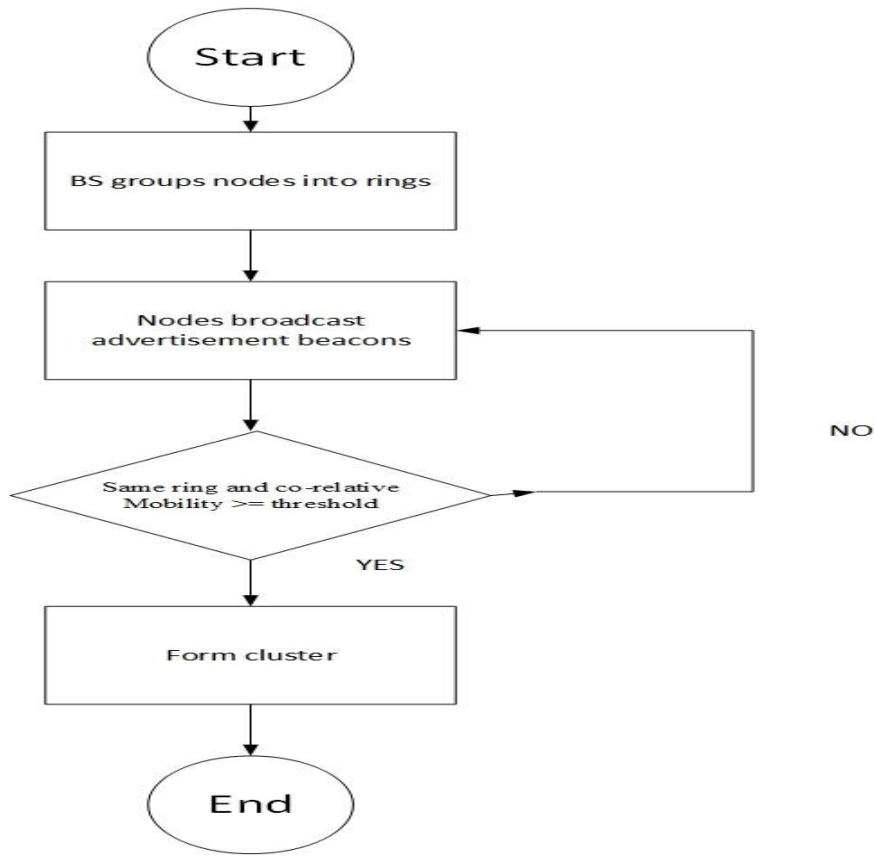

Fig. 2. 2 Logical flow of cluster formation scheme

## 2.5    Propsed Data Uploading Scheme

As already mentioned the data size generated by the IoT devices are relatively small; thus it may not require a high CQI; taking advantage of this trait IoT devices form a multi-hop data routing from CM to the BS (sink). Moreover only the CH(s) participates in the multi-hop data transmission also known as intercluster data transmission. Post cluster formation all the CMs upload their data to the CH using D2D links. A CH aggregates all the data uploaded by the CMs and start looking for next hop CH to relay the data in the adjacent ring closer to the BS.

The following steps summarizes the proposed data uploading scheme:

- The CM(s) uploads the data to the CH.
- The current CH aggregates the data and broadcasts the advertisement messages to find potential next hop CH in the adjacent ring closer to the BS.

- ▪ The interested potential relay nodes acknowledges the advertisement messages to the current CH.
- ▪ The current CH selects a next hop CH which has a highest co-relative mobility.
- ▪ The above mentioned steps will be repeated until all the data packets reach the BS.

The steps of the data uploading scheme is presented in the form of a flowchart in Fig. 2.3.

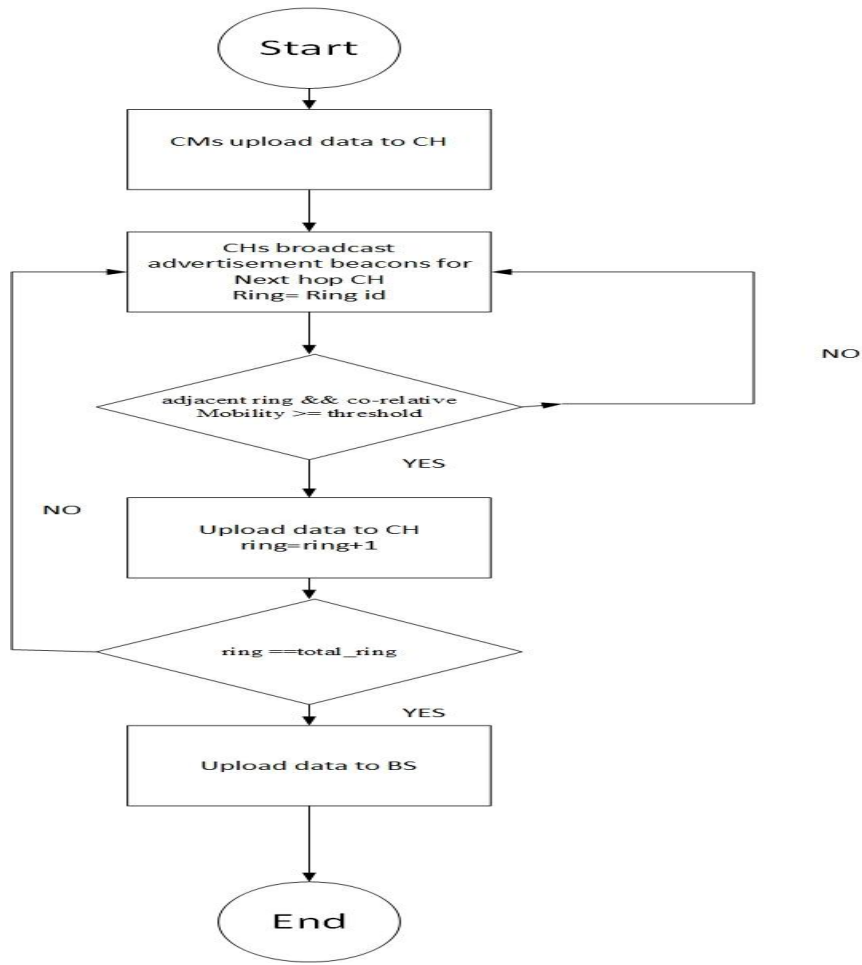

Fig. 2. 3 Logical flow of data uploading scheme

## 2.6    Implementation and Results

Our proposed D2D based IoT data aggregation scheme is simulated on Matlab 2012. To present the competitive performance of the proposal with other related schemes, we compared our findings with the findings of the recently published research, known as D2D-EE scheme [16]. The D2D-EE scheme is also based on D2D communications within the cluster of devices and aggregates the data in cluster heads before uploading it to BS, due to the similarity with our proposed scheme it is selected for comparison. During the simulation study, we have considered distance and velocity as random variables. Intuitively the proposed work is based on co-relative mobility; therefore, it is evident that in a massive IoT scenario velocity and relative distance of individual nodes have minimal effect on the overall performance. Hence, the claim that the proposed scheme includes both the stationary and non-stationary nodes is validated based on the theoretical analysis and we have not provided any plot to observe the effects of the mobility of nodes on the proposed system. Moreover, the experiments conducted during the simulation investigates the energy efficiency of the entire system in various scenarios. During the simulation study, we have developed a system that consists of N random nodes and a BS at the centre. Some parameters were kept constant while performing the experiments as shown in Table 2.1. The formation of clusters based on co-relative mobility is depicted in Fig. 2.4.

Table 2. 1 Simulation parameters of chapter 2

| Parameter | Value |
|---|---|
| Number of rounds procedure is repeated | 2 |
| $\alpha$ | 0.5 |
| $\beta$ | 0.5 |
| Beacon transmission range | 300m |
| Transmission time interval (TTI) | 1ms |
| Control packet length | 200 bits |
| Initial energy of each node | 0.5 joule |
| Maximum number of rings | 5 |

During the first experiment, we have plotted the energy efficiency (bits/joule) as a function of the number of IoT devices where the packet length is fixed at 10 bytes, as shown in Fig. 2.5. As depicted in the plot the energy efficiency is increasing with an increasing number of devices for both the schemes. Moreover, the proposed solution is highly superior in all the cases, and it is approximately five times more energy efficient than the D2D-EE scheme based on the number of participating devices.

During the second experiment, we have plotted the energy efficiency while varying the packet size from 0 to 100 bytes while keeping the number of devices constant at 50, as shown in Fig. 2.6. As evident from the plot the energy efficiency of our proposed solution is increasing till 40 bytes to attain the maximum energy efficiency and after that, it is aberrating from the pattern. In the case of the D2D-EE solution, it increases with an increase in packet size. Moreover, our proposed solution retained its supremacy over D2D-EE solution in nearly 80 % of the experimental outputs.

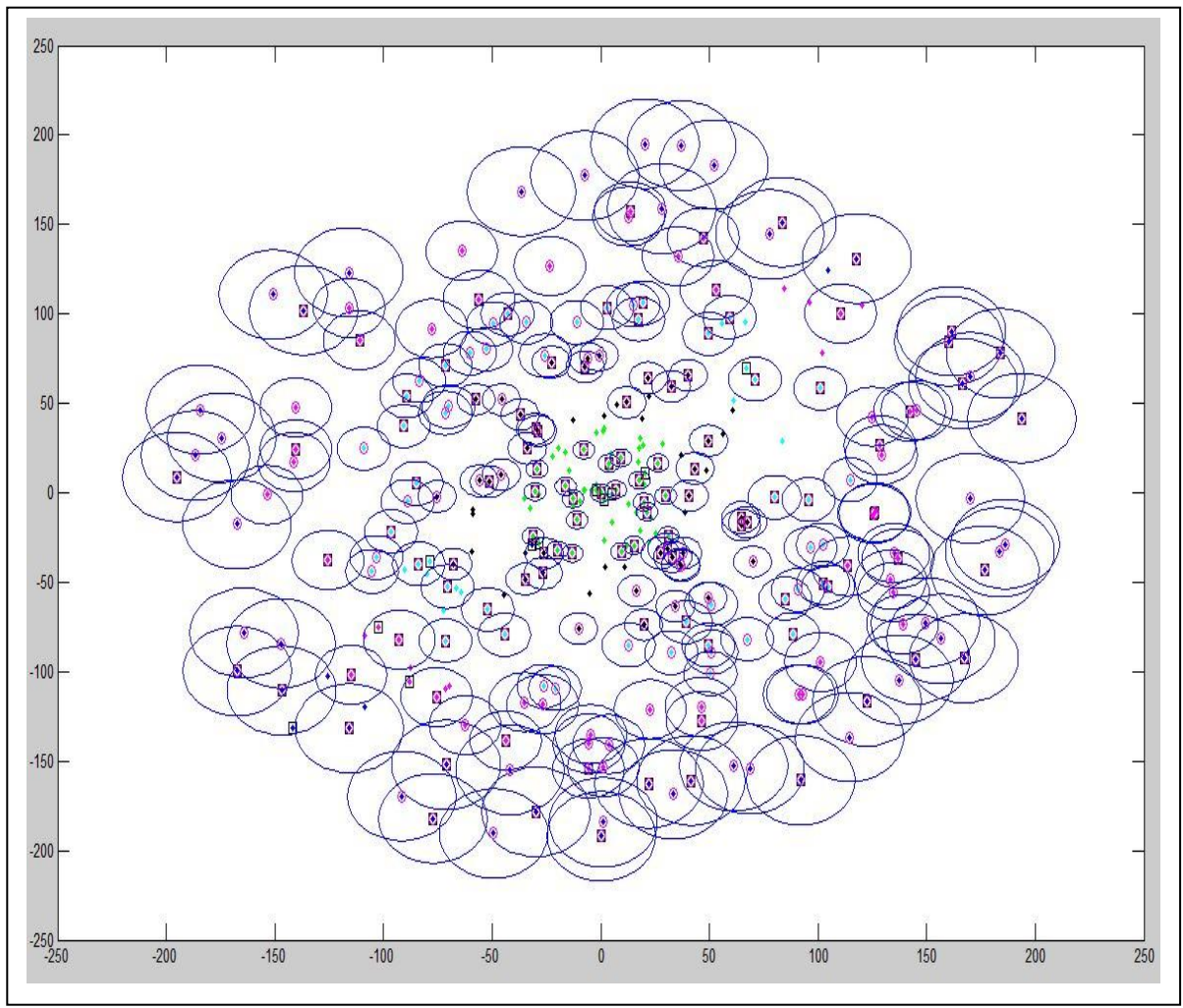

Fig. 2. 4 Formation of simulated clusters based on co-relative mobility

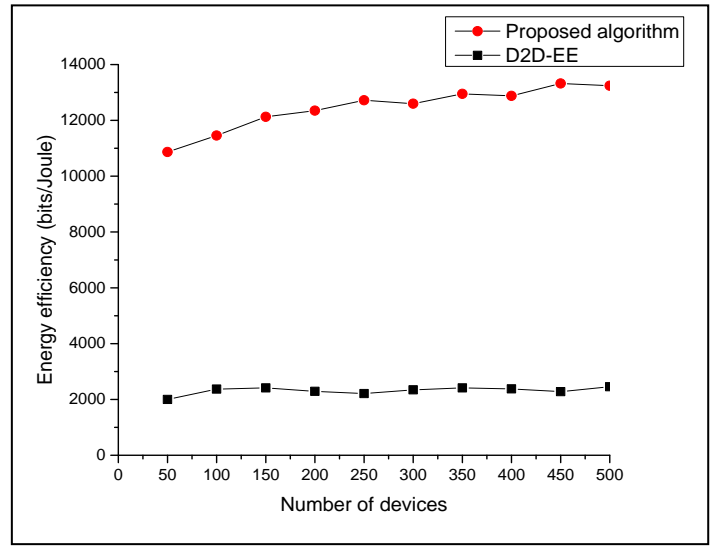

Fig. 2. 5 Energy efficiency vs Number of devices

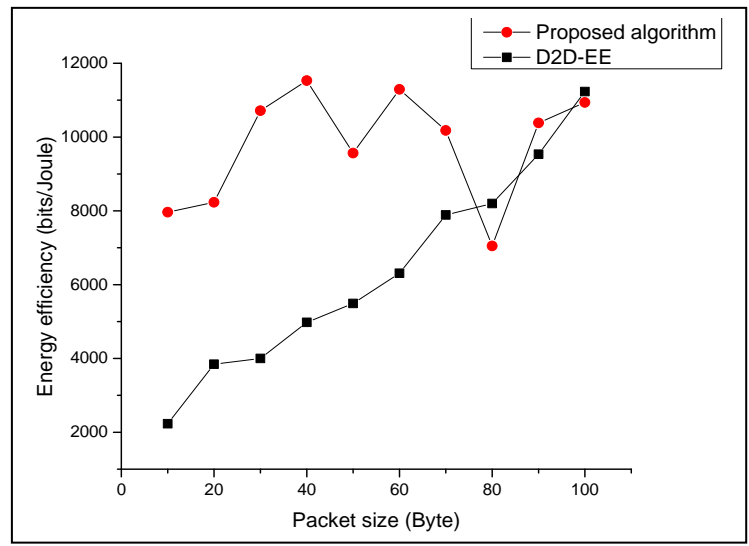

Fig. 2. 6 Energy efficiency vs Packet size

## 2.7    Chapter Conclusion

In this chapter, a cooperative IoT data aggregation scheme is proposed. The proposal consists of two main ideas which are cluster formation and multi-hop data routing schemes. This proposal can accommodate both stationary and non-stationary nodes. The cluster formation scheme uses a novel idea of co-relative mobility to group nodes into variable size clusters. The routing of data consists of inter-cluster multihop data relaying which adds to our main agenda that is energy efficiency. The simulation study shows that the proposal is highly energy efficient and can handle small as well as a large number of nodes. The proposal is suitable for scenarios where the solution is expected to be less computationally complex at the same time highly energy efficient for aggregation D2D links.

# Chapter 3 A Heuristic Solution to Data Veracity for Massive Raw IoT Data

## 3.1   Chapter Overview

The raw IoT sensor data collected using device to device communication (D2D) from unevenly distributed IoT sensor networks incur high uncertainties due to the presence of noise, outliers, missing readings and redundancy. These uncertainties if not treated may propagate throughout the system and degrade the overall performance, this is formally known as data veracity problem. This chapter considers the problem of data veracity in the raw IoT sensor data collected using D2D communication. The effectiveness of the research carried out by Balzano and Nowak in [21] and Hage and Kleinsteuber in [22] provide us strong motivation to address the problem of IoT data veracity problem in this dissertation.

Naive approaches such as embedding learning algorithm to process the data locally may destroy the global correlation and therefore adversely affect the decision making. Moreover, the so-called robust data analytics algorithms are actually compromising their decision accuracy for achieving high robustness against highly uncertain raw IoT sensor data. The proposed approach is based on the data aggregation architecture discussed in chapter 1 Fig. 1.5. This chapter discuses an application of the proposed acrrhitecture where it performs data aggregation to improve the quality of data by restoring the missing values, corrupted reading, outliers and redundancy without affecting the global intrinsic patterns of the raw data. The information content of the intrinsic pattern is higher than the rest of the data and can be used to estimate a better version (true data) of the data which is more reliable and can be directly utilized for data analytics application.

A big portion of existing literature is aimed at sensor data reduction, as a cure for uncertainty in sensor data. These approaches can be grouped as a) Random techniques such as sampling [23] and approximation [24]. These random techniques are computationally efficient, however, incur a high loss of important data that increases uncertainty. b) Supervised techniques such as regression [25] imposes high computational and communication costs. Regression also requires prior information that makes it impractical for implementations. c) Unsupervised techniques such as principal component analysis (PCA) cannot handle uncertainty. Other existing solutions [25] are too complex for practical implementations. Therefore there is an urgent need for a practical unified paradigm that can transform the highly uncertain raw IoT sensor data into reliable IoT sensor data.

This chapter advocates an alternative paradigm where the IoT data delivery mechanism and IoT data aggregation functionalities are made independent of the data analytics platform. The proposed paradigm enables the algorithm developers of the data analytics applications to completely focus on the accuracy of the decision making rather than the robustness of the application. The delivery mechanism is based on device to device communication similar to chapter 2, although, in this chapter the IoT devices are connected to the fog server using WiFi routers which acts as a base station. Furthermore the fog server also acts a data preprocessor for the aggregated raw IoT sensor data. This chapter proposes a novel data aggregation approach to improve the quality of the aggregated raw IoT data at the fog server. The proposed approach estimates a true sensor data matrix from the raw IoT sensor data. A true sensor data matrix is independent of uncertainties due to noise, outliers, missing values and redundant data. Moreover it also contains the intrinsic characteristics that truly represents the dynamics of the raw IoT sensor data and hence is more reliable than the raw IoT sensor data for data analytic applications. The proposed approach is divided into three steps: first a rough estimate of the original subspace is reconstructed using sample data further this subspace is optimized to track the dominant subspace; second, the dominant subspace is used to find the gain vector of the IoT devices; and third, the gain vector is used to estimate the true sensor data matrix. The approach is interdisciplinary in its core as subspace tracking [22] and blind equalization are extensively used in computer vision and signal processing, respectively.

The overall general contributions of this chapter are listed below:

- This chapter proposes a novel application of the proposed data analytics architecture discussed in chapter 1.
- It introduces an uncertainty model for IoT sensor data based on Shannon's entropy using an intuitive explanation. As presented in section 3.4.1.

- The reconstruction of data from the partial sets of raw data before the subspace tracking provides a major advantage over the similar approaches.
- This chapter presents a dominant subspace estimation method (section 3.5.1) which is robust to uncertainties of raw IoT sensor data. Moreover, the presented method does not require any prior information about the outliers or the missing values and can operate on fully and partially unobserved dataset.
- It presents a mathematical relationship between dominant subspace and true intrinsic sensor data (Eq. 3.10).
- It also presents an approach in section 3.5 to estimate a more reliable true sensor data which closely resembles the raw data but with nominal uncertainties.

This chapter is organized in the following fashion. Section 3.2 discusses the past notable literature using similar approaches to improve the quality of massive IoT data. Section 3.3 introduces the IoT data delivery and aggregation system architecture using D2D communication. Section 3.4 discusses the important assumptions, theoretical background and crucial information about the datasets. Section 3.5 serves the kernel idea of this chapter in two different subsections for two separate approaches, and also presents an algorithm for estimation of a more reliable sensor data from raw IoT sensor data. The section 3.6 deals with the simulation studies which includes experimental methodology, results and also discusses the limitation of the proposal. Finally, the chapter concludes by highlighting the major contributions and future works.

## 3.2    Related Work

This section discusses some of the notable approaches similar to the proposal that aims to clean the raw sensor data, in a chronological order and also compares them with our proposed approach. SMURF algorithm [26] is one of the first attempt to clean raw RFID data with window based adaptive smoothing filter. The statistical sampling theory is used to learn the window size continuously. Moreover different sampling approaches such as random sampling, systematic sampling, cluster sampling and quota sampling etc, are also used in the literature for raw sensor data cleaning. The major demerit of the sampling approaches is they generally preserves the local correlation and destroys the global intrinsic correlation of the raw data which increases randomness. Whereas our proposal aggregates the data in the form of a matrix and performs data cleaning based on the global view of the data.

The approximation approaches are frequently used for estimating a more reliable data from the raw data. Some notable approximation approaches are [27][28]. In general, the approximation approaches aggregates the data within an error bounded accuracy regardless of the actual system conditions. Approximation approaches have limited practical implementations in scenarios where accuracy is a critical requirement and also the complex nature of the theoretical analysis of the approximation approaches makes it unattractive to developers. Although our proposed solution requires estimation, however, the proposed approach has no error bounds on the accuracy.

Probabilistic approaches based on Bayesian statistics generate a model based on the user-defined features to distinguish important data within the raw data. [29] generates a model of the raw time series data of the mobile RFID using important features such as user mobility, object dynamics, corrupt readings. A particle filter is designed to track the desired information from the raw data. In [30] confidence interval is computed based on the multivariate normal probability distribution model of the IoT raw data stream to filter the undesirable data. This approach requires prior information about the data to generate a model. Our proposed approach requires no prior information and still generates satisfactory results.

The robust principal component analysis (RPCA) is the dominant paradigm to extract the desired subspace from the highly uncertain raw sensor readings. The outlier pursuit approach [31] of RPCA obtains the optimal low-rank matrix in the absence of noise. This approach is further extended in [32] as an iterative thresholding based approach with lower complexity than the former even in noisy environment. Some other notable approaches for the RPCA can be found in [33]. [25] presents a cluster-based data analysis approach using recursive principal component analysis to detect outliers in IoT data.  [31] [32]  [33] are limited to the subspace estimation, however, the proposed approach performs robust subspace estimation as the first step of our proposal and further uses the optimal subspace to generate reliable sensor data. In case of [25] which aims to only detect the outliers however the proposed approach aims detection and elimination of uncertainties.

## 3.3    System model

This chapter advocates the four-layer architecture (refer chapter 1) for massive IoT analytics. An example of massive IoT deployment can be found in [12]. In practical scenarios, the public and private clouds can be very far from the actual geographical location of the deployed sensor network. In case of the aforementioned example, the cloud servers are located in Beijing city, remote cloud servers may cause high latency and degrade the overall system performance. A fog server that is placed relatively closer to the sensor network deployment site can provide a platform for filtering and analysis sensor data close to the sensor network. This reduces the overall transmissions to the cloud thus improves the overall system performance. In this chapter, the fog acts as a data preprocessors and a base station that controls and monitors device to device communication among IoT devices. The fog server also aggregates the sensor data and generates a local sensor data traffic matrix, later this traffic matrix is processed using the proposed algorithm (section 3.5) to decrease the uncertainties caused by noisy readings, missing data, outliers and redundancy. The cloud server downloads the processed data from the fog servers and generates a central data traffic matrix for the centralized data analytics and decision making. The system model is depicted in Fig. 3.1.

All the IoT nodes are assumed to be connected with the fog server using Wi-Fi routers. The routing layer is also assumed to be anoutband device to device communication scenario where each IoT nodes $\{n_1....n\} \in N$ form clusters $\{l_1...l\} \in L$ with neighbouring nodes and every cluster has a cluster head. The chapter assumes that IoT nodes are stationary smart devices with limited power, storage and computational capabilities. Given the fact that IoT nodes are resource constrained, their role is limited to sensing and transmitting. The IoT sensor data vector at a particular time instant where each sensor is generating its reading can be represented as $y_{\{1...n\}} = [y_1...y_n]^T$. The fog server provides a unique id to each IoT device, and also segregates the IoT devices based on the serving cluster head. A cluster head is responsible for the aggregation of the data generated by the cluster members. A cluster head further relays the data to the next hop cluster head which is close to the fog node, therefore the data is delivered to the fog server using an inter-cluster multihop device to device communication.

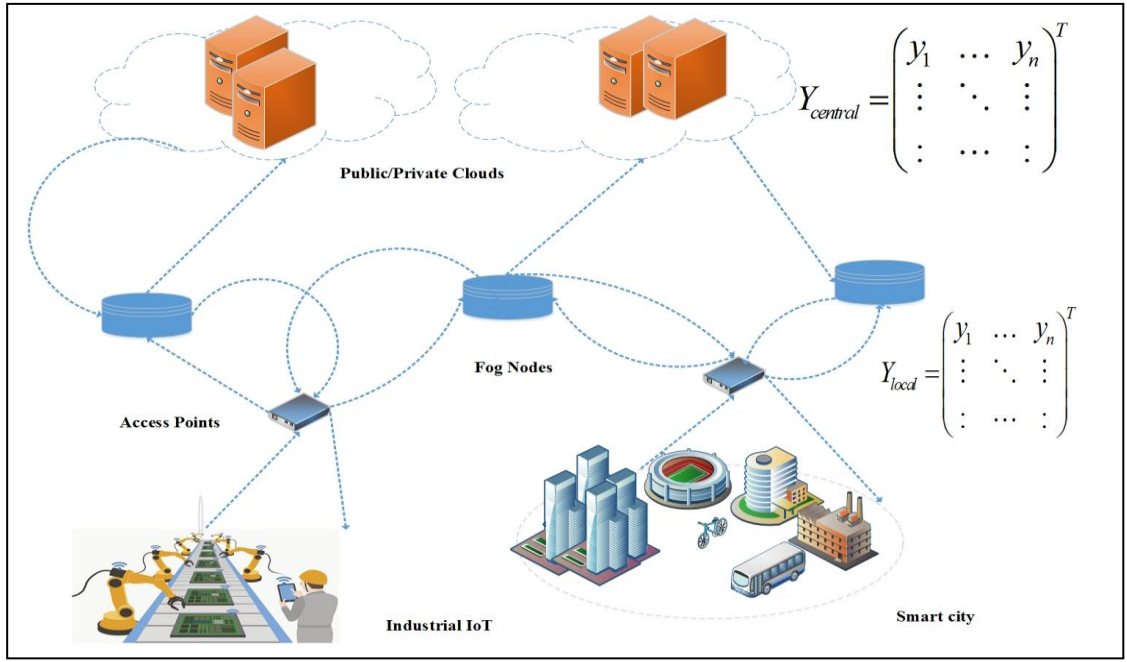

Fig 3. 1 Scenario of massive IoT data aggregation

The selection scheme of cluster head among the other cluster members and the selection scheme of a next hop cluster head for the inter-cluster device to device communication can be found in chapter 1.

## 3.4    Background

This section discusses some important prerequisites and notations (Table 3.1) that is extensively used throughout the upcoming sections of this chapter. This section introduces the uncertainty model and the characteristics of the data used in simulation studies. Finally this section present details of the D2D communication scenario and formulates the concept of true sensor data.

### 3.4.1  Uncertainty in IoT sensor data

Uncertainty is a broad term and still, there is no unified model that holds true for all the scenarios. Therefore, uncertainty in data is studied based on the specific scenarios. Existing literature [34] presents five different kinds of uncertainties namely:

Table 3. 1 Description of major symbols in chapter 3

| Symbol | Description |
|---|---|
| $n$ | Total number of IoT nodes |
| H(X) | The Shannon's entropy of a random variable X |
| $P(x_i)$ | The probability distribution of $x_i$ |
| $U$ | Basis of orthogonal dominant subspace |
| $\beta$ | Sensor offset values |
| $z$ | True sensor reading vector |
| $\alpha$ | Sensor gain vector |
| $Y$ | IoT sensor data traffic matrix |
| $\vec{Y}$ | $\vec{Y} = diag(Y)$ |
| $\overline{Y}$ | Zero mean centered IoT sensor data traffic matrix |
| $\widehat{Y}$ | Partial matrix based on the linear operator $\Phi(\overline{Y}) = \widehat{Y}$ |
| $L$ | Low rank approximation of $\overline{Y}$ |
| $h_\Delta$ | Smooth approximation of non convex penalty function h |
| $\theta$ | Complement of signal subspace $I - \delta$ |
| $\omega$ | Total no. of iterations |
| $\Delta$ | Smoothing parameter |

Shannon entropy, classification entropy, fuzziness, non-specificity and rough degree. The uncertainty discussed in this section is due to the presence of noise, outliers, missing readings and redundancy. The chapter perceives the IoT sensor data uncertainty based on Shannon's entropy model.

Shannon entropy $H(X)$: The chapter considers a random variable $X = \{x_1, \ldots, x_n\}$. The probability distribution of random variable can be represented as $P = \{p_1, \ldots p_n\}$.

$$H(X) = -\sum_{i=1}^{n} P(x_i) \log_2 P(x_i) \qquad (3.1)$$

Intuitively Shannon's entropy of a random variable is the amount of information, (i.e. the true subspace in this chapter) contained in the variable. This is not just the total number of different values for the random variable (i.e. the raw sensor data in our case). For example, the information in an email is not only the number of possible words or different usages of the words. Instead, the information of an email is proportional to the amount of surprises its reading causes. Based on the aforementioned explanation the chapter can safely infer that the information is not just the raw data, but it is a low dimensional pattern inside the raw data that carries most of the properties of the raw data. In section 3.5, the chapter uses this idea to design the data aggregation scheme.

## 3.4.2  D2D communication

3GPP has played a key role in establishing and optimizing the machine type communication (MTC) [35] in release 11, 12 and 13 that further evolves into D2D communication. The D2D communication can be classified in cellular D2D (inband) or unlicensed D2D (outband). The inband D2D can be further classified into underlay inband D2D where the D2D devices shares the cellular spectrum with other devices simultaneously and overlay inband D2D where the devices receive dedicated celluar spectrum. However this chapter focuses on outband D2D communications based on unlicensed spectrum using WiFi. This chapter computes the energy efficiency ($\eta$) of the D2D scenario with n number of IoT devices as:

$$\eta = \sum_{i=1}^{n} \frac{d_i}{E_i \cdot r_i \cdot TTI} \qquad (3.2)$$

Where $d_i$ is the total volume of data to be uploaded, $E_i$ is the average energy consumed to deliver a single packet, $r_i$ is total number of data packets to be uploaded by all the nodes and TTI=1 is transmission time interval which is constant to all packets.

### 3.4.3 Dataset

The proposal uses two different datasets, the first data set is a real sensor data set and the second one is a synthetic dataset. Both the datasets serve the purpose of experimental validation of two different aspects of the proposed solution, on one hand, the experiments with the real sensor dataset support the claim of real-world application and on the other hand, the Synthetic sensor data provide insights of the ideal scenario.

**Real world dataset:** The real world dataset comes from the deployment of IoT sensors inside a house located in Stambruges, Belgium [36]. The dataset consists of 4.5 months temperature and humidity monitoring data averaged over 10 minutes per transmission using wireless sensor networks. The Wireless sensor consists of a typical DHT 22 sensor for temperature and humidity measurements along with a Zigbee for radio transmission and an ATmega328P microcontroller embedded all in one module. The placement of sensors in the house can be seen in Fig. 3.2. Moreover, we have selected 400 independent data streams that have 400 sensor readings per stream and here we have considered each stream as an independent sensor. Furthermore, we have generated a 400 X 400 matrix of test data where we are considering 400 temperature readings from 400 different sensors.

Given the test data with zero mean principal component analysis (PCA) computes the axes with maximal variance and minimal redundancy this maximal variance also represents the dynamics of the raw IoT sensor data. Application of PCA to our real world sensor data shows that the major variance is captured by only a small portion of the dominant subspaces (Fig. 3.3). This observation provides us with a solid ground for the assumption that the raw IoT sensor data has an intrinsic lower dimensional subspace which carries the dynamics or important characteristics of the whole data.

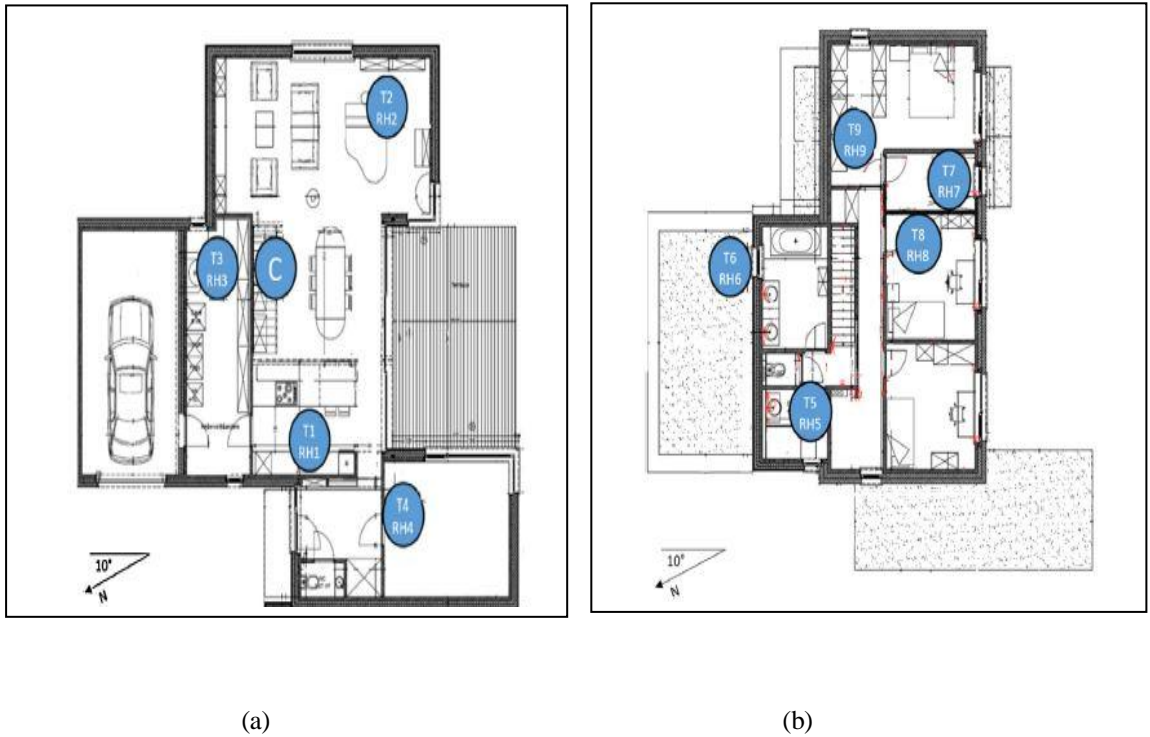

(a)                                                                        (b)

Fig 3. 2 Location of indoor temperature and humidity sensor deployment. (a) First floor (b) Second floor

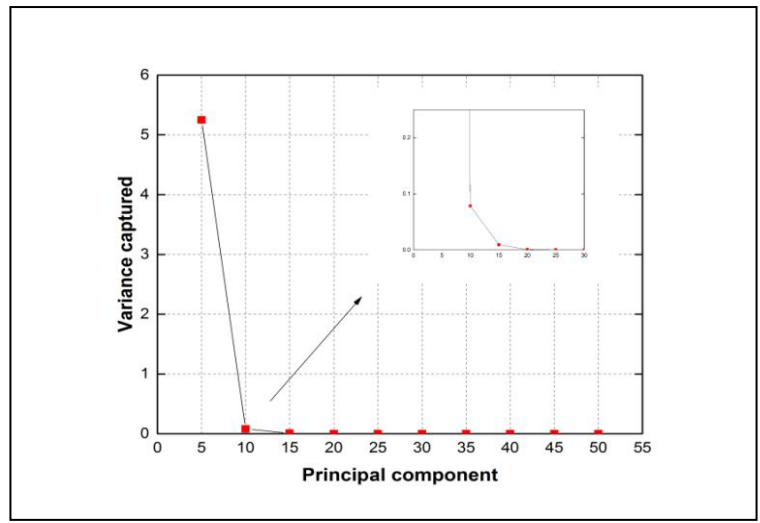

Fig 3. 3 Fraction of total link traffic variance captured by dominant subspace

**Synthetic Dataset:** We construct a 400 x 400 test data traffic matrix to emulate the data traffic of 400 IoT nodes, each generating 400 samples for a given time duration. Our test data traffic matrix is a sum of a fixed rank k matrix and a sparse matrix.

**Gain and offset:** The gain and the offset are generated using uniform distributions of [0.5, 1.5] and [-0.5, 0.5], respectively.

**Noise:** Here we generate a 400 x 400 matrix of pseudorandom mean zero Gaussian noise. In our experiments just to provide a visually simple relationship between noise variance and subspace perturbation we choose to directly perturb the subspace with noise rather than adding noise to the sensor data which eventually leads to subspace perturbation.

**Outliers:** we construct a 400 x 400 sparse matrix of outliers with a density of 0.2. The magnitude of the outlier is a uniform distribution of [-10, 10] and this magnitude is large enough to distort the computed subspace. The entries of the sparse outlier matrix follow Bernoulli distribution.

**Missing values:** We deliberately made 30 % of the IoT data traffic entries as missing values. Before adding perturbations to the data using noise, outliers and missing values a low dimensional rank k matrix is extracted from both the real world and synthetic data using singular value decomposition, after making all the singular values greater than k to zero ($\Sigma_{i \geq k} = 0$), then we multiply the factors ($U \Sigma V^T$). This low dimensional matrix serves as a true sensor data and is used to compare our results in section 3.6.

### 3.4.4  True Sensor Data Formulation

To maintain simplicity throughout the section the chapter assumes a univariate sensor scenario, but this example can be extended to multivariate sensor scenario as well, where a group of sensors are sensing different phenomenon. Here the chapter is making an important assumption that there lies a lower dimensional linear subspace which carries the true characteristics of the raw data, i.e. the true subspace of the n-dimensional Euclidean sensor space. Considering sensors as systems (Fig. 3.4), intuitively each sensor reading [21] can be represented as

$$y_{(i)} = \frac{z_{(i)} - \beta_{(i)}}{\alpha_{(i)}} \tag{3.3}$$

Where, $y$ is a vector of raw IoT sensor data, $\beta$ is a vector containing sensor offset values, $\alpha$ is a vector containing the sensor gains and $z$ is a vector containing the true values of sensor readings. Eq. 3.4 can be rearranged to form an equation that depicts the true sensor data out of raw IoT data traffic matrix.

$$z = \vec{Y}\alpha + \beta \tag{3.4}$$

Where $\vec{Y} = diag(Y)$ and each non zero component of $\vec{Y}$ is the mean of $y_{\{1...n\}}$ at a given time instant. Sensors react to changing physical conditions i.e. the stimuli by alternating the electrical properties such as resistivity, voltage and current. The offset is generally used to balance the bias and is explicitly mentioned on the datasheet of the sensor [37]. Therefore it is practical to assume that each IoT node is aware of its offset value.

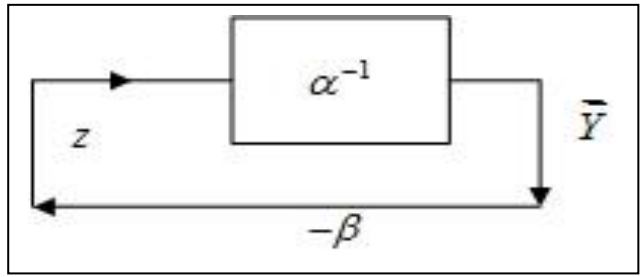

Fig 3. 4 Fraction of Total link traffic variance captured by dominant subspace

## 3.5    Main approach

As discussed in section 3.4.4 the chapter assumes that the raw IoT sensor data has an intrinsic true subspace which carries most of the dynamics of raw data and in section 3.4.1 we have defined the information content i.e the true data as a more reliable data with minimal or no uncertainties; also fit for data analytics. Therefore to solve the IoT data veracity problem this section seeks an approach that finds a non redundant linear intrinsic subspace even in the face of high uncertainties, then this subspace is used to generate more reliable data which emulates the true dynamics of the raw IoT data with no or minimal uncertainties. This section is divided into two parts the subsection 3.5.1 discusses the approach to estimate the subspace and in subsection 3.5.2 discusses the approach to the generate true IoT sensor data.

### 3.5.1  Robust Subspace Estimation

The robust subspace estimation focuses on two steps first reconstruction of subspace using sample data and second iteratively tracking down the intrinsic dominant subspace. $\overline{Y}$ is the zero mean centered $m \times n$ IoT sensor data traffic matrix (Fig. 3.1). The dominant subspace (DS) can be defined as $k$ uncorrelated linear components of $\overline{Y}$ as per Eq. 3. 5.

$$DS(\overline{Y}) = U_i^T \overline{Y} \tag{3.5}$$

Where $k \leq m$ and $\{U_i\}_{i=1}^k$ are the k orthogonal eigenvectors of $\Sigma_{\overline{Y}} = \mathrm{E}\left[\overline{Y} \times \overline{Y}^T\right]$. Moreover $\{U_i\}_{i=1}^k$ is also the first $k$ columns of left singular vector $U$ in the singular value decomposition of $\overline{Y} = U \sum V^T$. The chapter assumes the traffic matrix data model as $\overline{Y} = L + S$, where $L$ is a low rank approximation and $S$ is the sparse component of $\overline{Y}$. Since the rank of $L$ is less than $k$, $L$ matrix can be written as $L = U\overline{Y}$. Initially this approach chooses a partial matrix $\widehat{Y}$ as $\widehat{Y} \subseteq \overline{Y}$ using a linear operator $\widehat{Y} = \Phi(\overline{Y})$. This partial matrix is iteratively reconstructed. Finally the goal is to find the optimal basis for the intrinsic subspace using the partial entries (reconstructed matrix) of the IoT data matrix i.e the solution to Eq. 3. 6.

$$\min_{rk(L) \leq k} \left\| \widehat{Y} - \Phi(L) \right\|_0 \tag{3.6}$$

Eq. 3.6 is computationally complex and conventional optimization techniques is too slow for practical implementation, therefore the approach introduces a non-convex sparse penalty function $h$  Eq. 3. 7.

$$\min_{rk(L) \leq k} h\left(\widehat{Y} - \Phi(L)\right) \tag{3.7}$$

The approach considers $h_\Delta$ as a smooth approximation of $h$ which can be formulated as $l_P$ norm,

$$h_\Delta(\widehat{Y}) \approx \sum_{j=1}^n \sum_{i=1}^m (\widehat{y}_{ij}^2 + \Delta)^{p/2}, 0 < p < 1 \tag{3.8}$$

The parameter $\Delta$ can be tuned to remove outliers as well as induce sparsity in the outliers. This smoothing penalty function enjoys the advantages of $l_0$ regularization with faster convergence. Based on empirical information [22] it is observed that relatively large delta leads to faster convergence and relatively small delta leads to sparse output. Finally this approach is looking for an orthonormal basis for the dominant subspace as shown in the following equation.

$$U^{(i+1)} = \underset{U^T U = I}{\arg\min} \, h_\Delta(\widehat{Y} - \Phi(UU^T L^{(i)})) \tag{3.9}$$

The above optimization problem (Eq. 3.9) is solved iteratively using alternating direction method of multipliers model [38]. To achieve a smooth $U$ throughout, $\Delta$ is shrinking iteratively with each step. Moreover, the resultant U is not uniquely determined. Since the approach is working with $\widehat{Y}$ (Eq. 3.9) which is a subset of the raw data traffic and not the complete raw data itself, the dominant subspace is resistant to various uncertainties embedded in raw IoT sensor data.

## 3.5.2 Estimation of True Sensor Data

Based on the Eq. 3.3 and Eq. 3.4 of section 3.4.4 the chapter assumes that there lies a linear subspace inside the unstructured massive raw IoT data. In this section, the chapter further discusses an approach to extract the true sensor readings out of massive raw IoT data. This approach consists of two steps, first the approach projects the incoming raw IoT sensor data to the dominant subspace and compute the sensor gain vector and secondly the approach estimates the true IoT sensor sensor matrix based on the gain vector.

The approach considers the signal space as $\delta = UU^T$ and $\theta$ be the complement of the signal subspace. Then from Eq. 3.4 the approach can safely assume that every point in $z$ also belongs to $\overline{Y}$ i.e. $z \subseteq \overline{Y}$. The approach observes a relation between the dominant subspace basis and the true intrinsic subspace as shown (Eq. 3.10),

$$\theta z = \theta(\vec{Y}\alpha + \beta) = 0 \tag{3.10}$$

Theoretically this relation holds since all the column vectors of $\theta$ are orthogonal vectors and $z^T \subseteq \delta$ is also orthogonal vector. Moreover as discussed in section 3.4.4 the offset $\beta$ is a constant. Therefore the approach is only interested in recovery of the gain vector $\alpha$ from Eq. 3.10. This section is now about to derive an equation for mean centered traffic from Eq. 3.10 only dependent on $\alpha$.

$$\theta \left( \left( \frac{1}{n} \sum_{i=1}^{n} Y \right) \alpha + \beta \right) = 0 \tag{3.11}$$

Rearranging Eq. 3.11 as follows

$$\theta \left( \frac{1}{n} \sum_{i=1}^{n} Y \right) \alpha = -\theta \beta \tag{3.12}$$

Substituting Eq. 3.12 in Eq. 3.10

$$\theta \left( \bar{Y} - \left( \frac{1}{n} \sum_{i=1}^{n} Y \right) \right) \alpha = 0 \tag{3.13}$$

$\bar{Y}$ is a zero mean centered IoT data traffic matrix. Therefore Eq. 3.13 can be written as follows

$$\theta \bar{Y} \alpha = 0 \tag{3.14}$$

Moreover in the presence of even minor uncertainties it is not possible to solve Eq. 3.14 for $\alpha$ in a closed form fashion. Hence this approach looks for a robust optimal solution rather than the true solution for Eq. 3.14, hence this approach models Eq. 3.14 as an optimization problem.

$$\arg \min_{\alpha} \left\| \theta \bar{Y} \alpha \right\|_2^2 \tag{3.15}$$

A reasonable equivalent for the optimization problem provided in Eq. 3.15 is to find right singular vectors of $\theta \bar{Y}$ associated with singular values in ascending order. Once the approach finds an optimal set of $\alpha$, using Eq. 3.4 one can find the true intrinsic subspace present inside the raw IoT big data. The steps for data aggregation are summarized in algorithm 1. This algorithm starts with a SVD of an arbitrary part of the whole IoT sensor data traffic matrix $\overline{Y_0}$ and generates a rough estimate of $L$. Initially the $\Delta$ is relatively large which enforces fast subspace reconstruction until a good rough estimate of the subspace is achieved later the $\Delta$ is reduced iteratively to further optimize the subspace. The resultant dominant subspace is used in step 2 and an optimal gain vector is computed. Finally in step 3 a more reliable true sensor data is generated using Eq. 3.4.

---

**Algorithm 1：Data Aggregation Scheme**

---

**Initialization:** $\Phi(\overline{Y_0}) = \hat{Y}$, perform $SVD(\overline{Y_0})$ to obtain $U^{(0)}$ and

$$L^{(0)} = U^{(0)}U^{(0)T}\overline{Y_0}$$

**Set：** $\Delta^{(0)}$, $\Delta^{(\omega)}$ and $\Omega = \left(\Delta^{(\omega)} \middle/ \Delta^{(0)}\right)^{1/(\omega-1)}$

---

1: **for** i=1: $\omega$; **do**

2:      find optimal $U^{(i+1)}$ {Step 1}

3:  $U^{(i+1)} = \underset{U^T U = I}{\arg\min} \, h_\Omega(\hat{Y} - \Phi(UU^T L^{(i)}))$

4:       $L^{(i+1)} = U^{(i+1)}U^{(i+1)T}\overline{Y}^{(i+1)}$

5:       $\Delta^{(i+1)} = \Omega\Delta^{(i)}$

6:      **end for**

7:   retain only k rows of U

8: find optimal $\alpha$        {Step 2}

9:  $\alpha^* = \underset{\alpha}{\arg\min} \left\|\theta\overline{Y}\alpha\right\|_2^2$

10:  $z = \vec{Y}\alpha + \beta$           {Step 3}

---

## 3.6    Performance Evaluation

The first part of this section discusses the experimental setup and the second part discusses the experimental results.

### 3.6.1  Methodology

The performance evaluation is focused on answering these crucial questions: What are the characteristics of the subspace? How accurately the proposed approach estimates the true IoT sensor data in the face of high uncertainties? Is the proposed scheme scalable for IoT applications?

To address the first issue, we present a graphical relationship between noise variance and subspace error. To answer the second question we present the estimated true IoT sensor data in the presence of high Gaussian noise with outliers and missing values. Here we also compare our approach of true data estimation using robust subspace estimator against a baseline algorithm. We proceed to answer the third question of scalability by computing the energy efficiency while varying the number of devices based on the D2D network architecture.

Our approach is focused on improving the quality of massive IoT sensor data, hence we believe the prowess of the approach can be proved even with a small volume of data.

**Baseline:** Principal component analysis (PCA) is a classical tool for low dimensional linear subspace approximation called the principal components. The effectiveness of singular value decomposition (SVD) plays a big role towards the popularity of PCA. However, SVD based PCA is fragile to high uncertainty and may generate arbitrary subspaces. This chapter selects PCA as a baseline since PCA is very popular and also has many variants, therefore may serve as a good baseline algorithm for the proposed algorithm. Moreover, it is important to note that the baseline algorithm uses PCA only to compute the dominant subspace i.e. step 1 for algorithm 1 and the dominant subspace is used to generate true sensor data as discussed in step 2 and 3 of algorithm 1.

### 3.6.2 Experimental Results

To demonstrate the efficacy of our approach we have performed four experiments, first we plot the subspace reconstruction error (Eq. 3.16) as a function of noise variance to illustrate the relationship between subspace reconstruction error with the noise variance (Fig. 3.5). We consider the rank of low dimensional subspace k=10 for all the experiments. Based on Fig. 3.5 the noise variance and the subspace reconstruction error have approximately linear relationship i.e. the noise variance is directly proportional to the subspace reconstruction error. Typically, this relationship depends on the nature of the noise, moreover, we have assumed that adding noise will increase the error (subspace error in our scenario). Since now we know the relationship between subspace error and noise variance we use subspace reconstruction error as a reference feature to present varying levels of subspace perturbations based on the noise variance. The reader should note that the noise variance ranges from [-0.01, 0.06] to keep the perturbed data correlated and if we increase the noise variance far more than this range it would be considered as an outlier that already depicted using experiment 4. We compute the subspace reconstruction error as follows:

$$\frac{1}{n}\sum_{i=1}^{n}\frac{\left\|U^{T}\tilde{Y}-U^{T}\overline{Y}_{i}\right\|_{2}}{\left\|U^{T}\overline{Y}_{i}\right\|_{2}} \qquad (3.16)$$

Where $\tilde{Y}$ is the estimated mean sensor data matrix and $\overline{Y}$ is the IoT sensor data traffic matrix. We compute the true sensor data estimation error as follows:

$$\frac{1}{n}\sum_{i=1}^{n}\frac{\left\|\tilde{z}_{i}-z_{i}\right\|_{2}}{\left\|z_{i}\right\|_{2}} \qquad (3.17)$$

Where $\tilde{z}$ a vector of the estimated true IoT sensor data of node $i$ and $z$ is a vector of the true IoT sensor data of IoT node $i$.

For the next experiment we adulterate the subspace with Gaussian noise. In this experiment just to provide a visually simple relationship between noise variance and subspace perturbation we choose to directly perturb the subspace with noise rather than adding noise to the sensor data which eventually leads to subspace perturbation. As shown in the Fig. 3.6, we have successfully estimated the true sensor data to a reasonable extent for both the datasets.

For third experiment we further added high Gaussian noise along with considerable amount of outliers to our datasets and estimated the true sensor values. Fig. 3.7, shows that our algorithm estimates true IoT sensor data to a reasonable extent even in the presence of big outliers. Moreover we also compare the proposed approach with the baseline approach where we compute the subspace using classical PCA and then estimate the true sensor data same as discussed in algorithm 1. It can be seen that the baseline approach generates very high true sensor data estimation error even with low reconstruction error (low noise variance) and also generates random results without any pattern. Moreover it is well known from the empirical studies [31] [32] that SVD based PCA is fragile to high uncertainties and generate random subspace basis.

For the fourth experiment we add high Gaussian noise along with synthetic missing values to the dataset. Fig. 3.8, shows that our algorithm can estimate true sensor data to a reasonable extent even in the presence missing values. Again SVD based PCA generates high estimation error along with random results. In general, it can be observed that the accuracy of our approach on real world sensor dataset is lesser than the synthetic dataset, it is due to the fact that the sensor dataset is already corrupted with noise other than the synthetic additive gaussian noise.

Apart from the above four experiments aimed to validate the core data aggregation approach we also check the scalability of the D2D communications for its practical implementation (Fig. 3.9). Here we simulate the energy efficiency (Eq. 3.2) while varying the number of IoT devices based on the system model in section 3.3. We compare the presented cluster based D2D communications for IoT data delivery with recent D2D-EE scheme [16] and the non cooperative standard LTE-A solution.

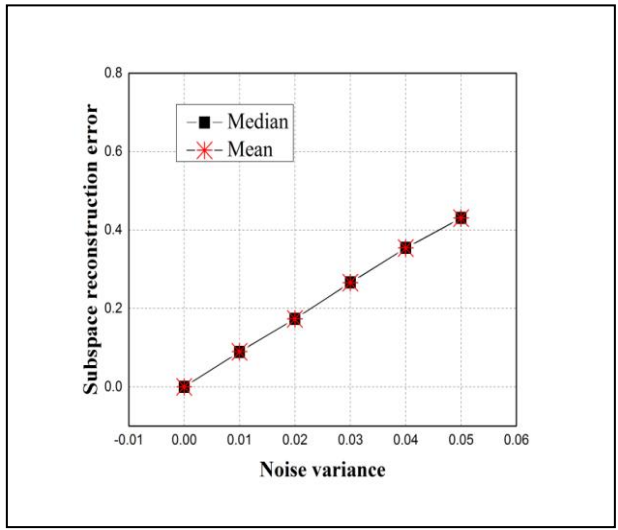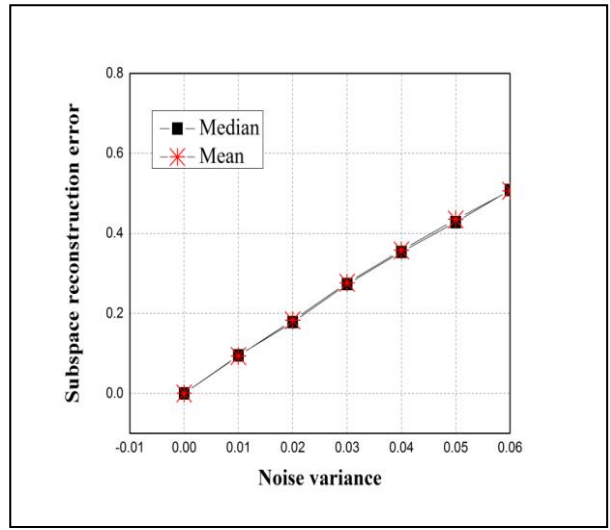

(a)                                                                    (b)

Fig 3. 5 Subspace reconstruction error as a function of Gaussian noise variance. (a) Synthetic dataset. (b) Sensor dataset

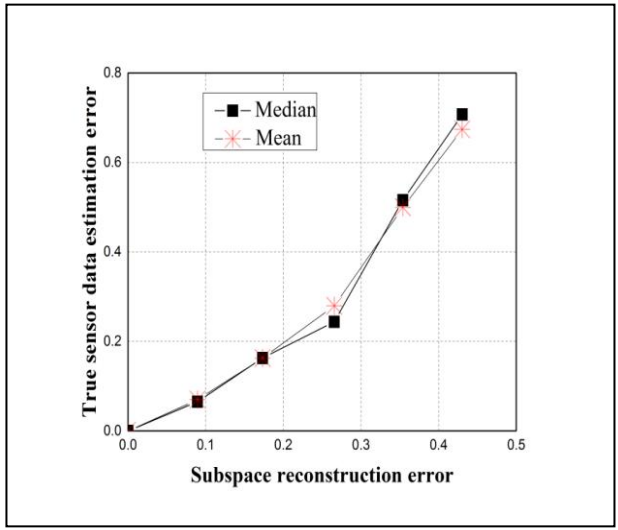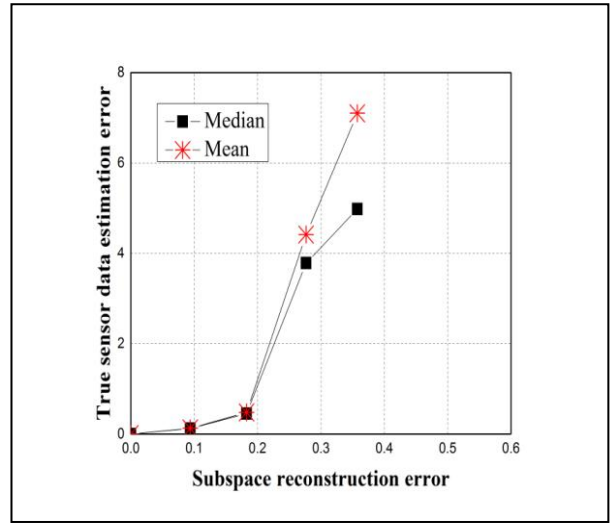

(a)                                                                    (b)

Fig 3. 6 True sensor data estimation error as a function of subspace reconstruction error. All the values are mean over 10 samples, where n=400 and k=10. (a) Synthetic dataset. (b) Sensor dataset

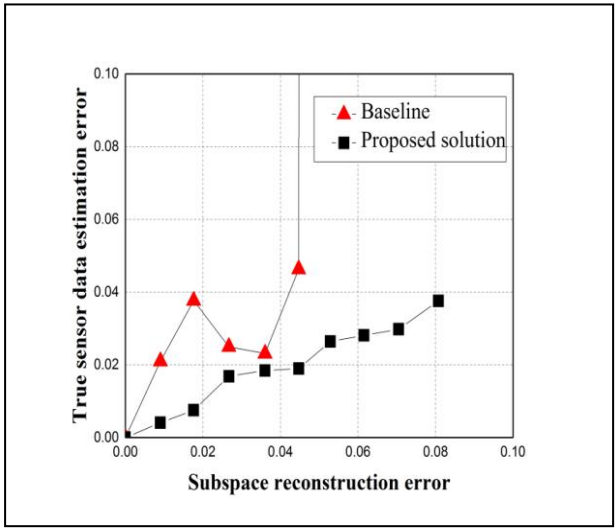 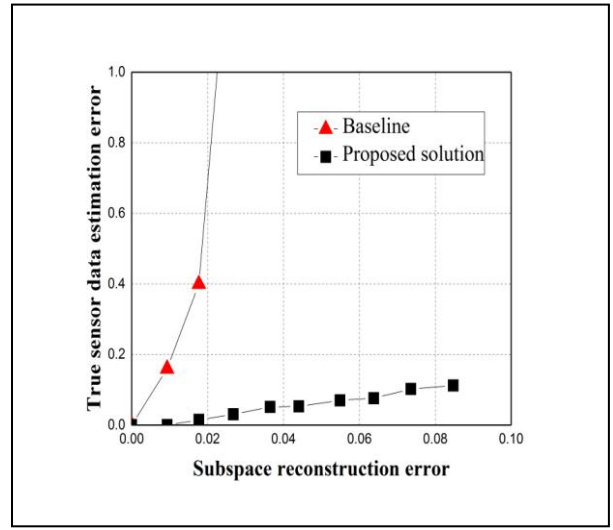

<div align="center">(a)                                  (b)</div>

Fig 3. 7 True sensor data estimation error asa function of subspace reconstruction error in the presence of outliers. Size of outliers between [-10, 10] with a density of 0.2 and all values are mean over 10 samples, where n=400 and k=10. (a) Synthetic dataset (b) Sensor Dataset

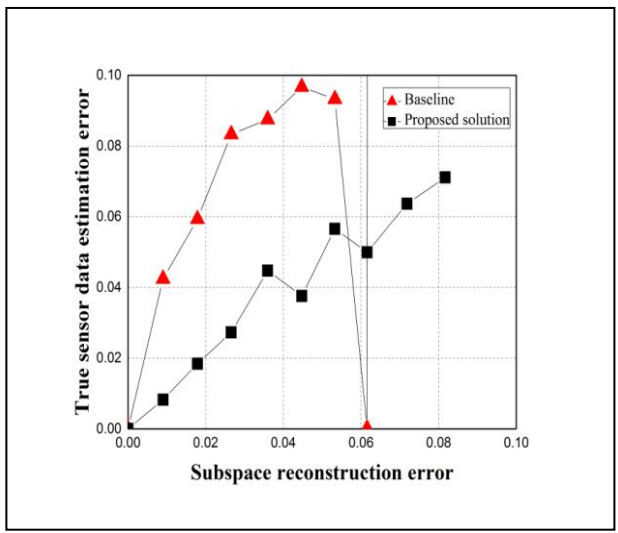 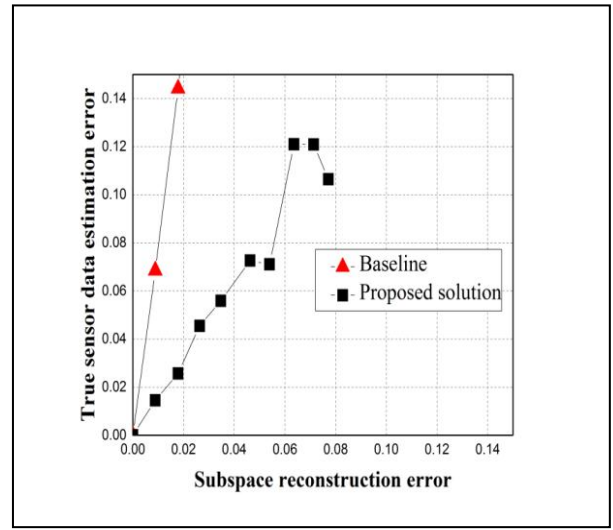

<div align="center">(a)                                  (b)</div>

Fig 3. 8 True sensor data estimation error as a function of subspace reconstruction error in the presence of missing value. 30 % missing values and all values are mean over 10 samples, where n=400 and k=10. (a) Synthetic dataset (b) Sensor dataset

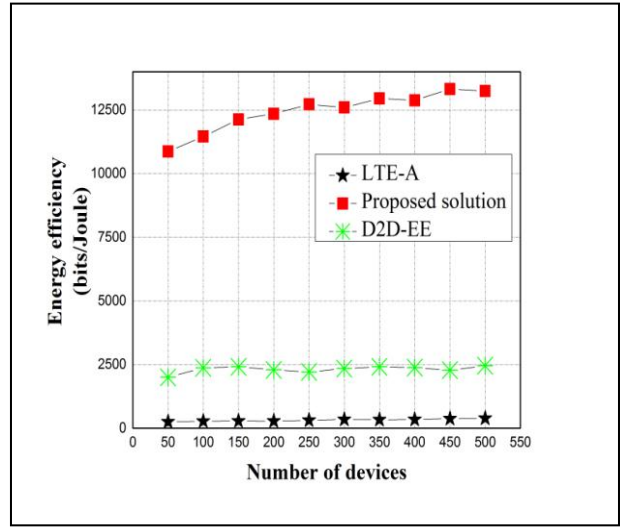

Fig 3. 9 Scalability, the packet size is 10 bytes

### 3.6.3 Discussion

The baseline approach is fragile to high uncertainties due to the fact that the single value decomposition approach directly processes raw IoT sensor data to compute the optimal subspace basis. The empirical observations [31] [32] show that it rather generates random subspace basis in the presence of high uncertainties such as big outliers and missing values. Therefore due to the selection of a random subspace basis instead of the optimal one the baseline algorithm fails to estimate true sensor data and generates random results. The proposed approach does not compute the optimal subspace directly using the raw IoT sensor data, instead at the beginning it roughly reconstructs the subspace using a partial matrix and then iteratively tracks down optimal dominant subspace. Moreover using the optimal subspace the proposed solution efficiently estimates true IoT sensor data even in the presence of high uncertainties. Therefore the proposed approach is robust to high uncertainties and performs better than the baseline approach.

**Limitation:** The proposal makes a common yet stringent assumption that each basis is orthogonal to the previous one, moreover, if the data is arranged in non-orthogonal basis then the dominant subspace estimation approach (section 3.5.1) generates random outputs. The proposal also assumes that the outliers are bigger in size than the inliers, moreover, if both outliers and inliers are comparable in size then the proposed scheme may fail to serve the purpose.

## 3.7    Chapter Conclusion

This chapter deals with the data veracity problem due to the presence of noise, outliers, missing readings and redundancy in the raw IoT sensor data. This chapter presents a data aggregation approach to efficiently tackle the data veracity problem. The data aggregation generally consists of a data delivery mechanism and a data preprocessing algorithm. The data delivery mechanism is designed with clustering based D2D communication among participating IoT devices. The data preprocessesing approach which is the kernel of this chapter is an amalgamation of two independent approaches i.e. robust dominant subspace estimation and tracking and true sensor data estimation. In robust subspace estimation, the proposed algorithm first reconstructs a rough estimate of the subspace using sample raw IoT sensor data and further optimizes the roughly estimated subspace to track the dominant subspace. This subspace is used to generate the reliable true sensor data matrix. As presented in the simulation studies the proposed algorithm can estimate true sensor data in the presence of noise, outliers, missing values and redundancy.

# Chapter 4 Federated Filtering for IoT Medical Data Aggregation

## 4.1    Chapter Overview

World Health Organization (WHO) recently reports [39] a global health workforce shortage of 12.9 million during the coming decade. This expected shortage accompanied by various other factors have inspired a slow but steady paradigm shift from conventional healthcare to the smart healthcare [40]. The smart healthcare enables patients with round the clock monitoring and feedback and is also expected to automate critical operations inside ICU. Internet of Things (IoT) is widely accepted [41] as a crucial driver to the connected healthcare paradigm.

A typical wearable IoMT device consists of a tiny battery which in most cases is nonchargeable [13], and this leads to disposal of the equipment once it is out of charge. A significant cause of speedier disposition of IoMT devices is due to the dominant cloud computing paradigm of pushing all the collected data to the distant cloud servers for analytics and decision making. This phenomenon incurs a significant loss of power due to high communication overhead. Moreover, it also exposes the aggregated sensitive medical data to the security risks. This chapter considers the problem of high power loss, exposure of medical data privacy and high latency in cloud based healthcare analytics. It is an interesting problem as it has social implications; moreover, governments and industries, are investing a lot of money and resources to develop a future healthcare infrastructure.

This chapter presents an algorithmic framework namely Federated Filtering Framework (FFF) (Fig. 4.1) for IoMT supported by theoretical analysis. The proposed framework presents an alternate solution to the issues of energy efficiency, latency and privacy for resource-constrained IoMT devices. In brief, each IoMT device computes a local model of the data and shares this model with the fog server. The fog server's role is threefold. First, it predicts a data matrix (aggregated data matrix) using aggregated model average (Section 4.5); second, it computes and delivers filter parameters for all the IoMT devices and finally performs decision making using the aggregated data matrix. To control the eigenvalue perturbation of the data matrix that compromises the decision accuracy this chapter derives a theoretical relation between the local filtering parameter with the global tolerable eigenvalue perturbation using Matrix Perturbation Theory (MPT).

Overall, the contributions of the chapter are as follows:

- a theoretical relationship between local time series filtering and perturbation error of aggregated data matrix.
- the implementation of federated decision making framework using filters.
- a lightweight fully unsupervised local subroutine (algorithm 1).
- the filter model averaging (algorithm 2) that preserves the privacy and demands few updates.
- a practical framework for IoMT data aggregation.

This chapter is organised in the following fashion. Section 4.2 discusses the related work. Section 4.3 presents the system model. Section 4.4 presents the theoretical analysis. Section 4.5 presents the kernel of the chapter which is Federated Filtering Framework. Section 4.6 shows the experimental evaluation, and finally, the chapter concludes by highlighting the significant contributions and future work.

## 4.2    Related Work

This section compares the proposed framework with three closely related genres of research that includes IoT in healthcare, prediction based IoT systems and federated learning approaches in networks.

### 4.2.1  IoT in Healthcare

The dominant paradigm for IoT based healthcare analytic systems [41] can be categorized as cloud computing-based health monitoring and mobile computing based health monitoring. Both the scenarios mentioned above very frequently push data to the server (cloud server/mobile device) for decision making. This chapter is firmly against the continuous transmission of data and presents a prediction based data aggregation scheme with error bounds to ensure the fidelity of the decision making.

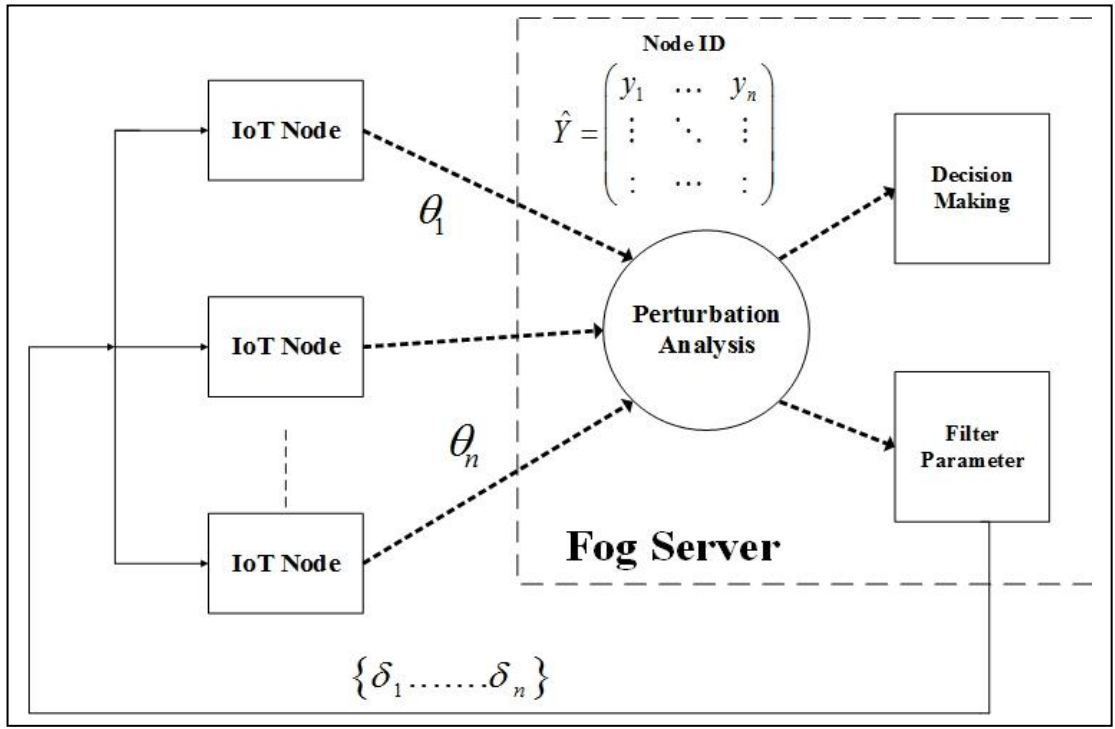

Fig. 4. 1 Federated Filtering Framework

Some recent use cases of IoT based healthcare analytics such as [42] [43] also advocates centralized decision making, however, both of them lacks a theoretical formulation to ensure decision-making accuracy.

### 4.2.2  Prediction based IoT systems

The literature [44] reports several prediction based approaches for reducing the communication overhead in sensor networks. The prediction [45] based approaches are categorized into single prediction approaches and dual prediction approaches. In the case of single prediction approaches the system performs prediction in only one location whereas in the case of dual prediction approaches the system performs prediction at a local node along with the central server. Some notable prediction schemes applicable for both the categories mentioned above are adaptive filtering scheme [46], Autoregressive filter, Autoregressive Integrated Moving Average filter (ARIMA) [45], Kalman filtering and machine learning techniques. Although some of the prior approaches can provide better accuracy for the model generation at the IoT device however given the severe computational constraints of the IoMT devices these approaches are not practical for local processing. Moreover, none of the earlier approaches shows any relationship between local and global processing using theoretical upper bounds.

### 4.2.3  Federated Learning in Networks

The original federated learbning scehme [47] is a prospective candidate for solving latency and privacy issues, however since it is originally developed for mobile devices the scheme assumes that the distributed client devices are resource abundant. Hence deferated learning paradigm advocates implementation of deep learning algorithms inside distributed client devices for local model generation. The second major demerit incurred by the federated learning paradigm is the computation of local model and the current global model are fully isolated hence it is very difficult to centrally regulate the formation of a desirable local model.

The effectiveness of federated averaging algorithm for distributed training proposed by Mcmahan et al. [47] provides strong motivation to develop a federated filtering framework for IoMT devices. Moreover, there are also other notable distributed optimization approaches [48] that improves communication efficiency. All the distributed and federated approaches in the literature are highly complex to run in a tiny IoMT device furthermore, they aim to perform decision making at the user device. The proposed Federated Filtering Framework on the other hand proposes a very lightweight subroutine for severely resource constrained IoMT device and also aims to perform decision making at the server using local shared model. The proposed framwoek also proposes a theoretical framwoek from regulating the local model generation based on the current global model. In a nutshell our work in chapter 4 inherits all the positives of the federated learning paradigm and tweaked the paradigm to suit the resource constrained IoMT scenario.

### 4.3    System Model

The system model considers a massive IoMT scenario where n number of IoMT devices are cumulatively working towards sensing a particular phenomenon. All the IoMT devices are connected to the fog server(s) using Wi-Fi links. Each IoMT device $\{N_1, \ldots, N_n\} \in N_i$ generates a time series data stream. This chapter assumes a centrally aggregated matrix $Y$ also known as global matrix (real) of size $m \times n$ where each column ($Y_i$) represents a particular IoMT device, and each row has a sensor reading of every 30 seconds. This generation of a global matrix $Y$ requires continuous transmission of data to the fog server. However, this chapter doesn't advocate a continuous push and therefore proposes a prediction based framework. The fog server generates an aggregated data matrix ($\hat{Y}$); i.e. a predicted data matrix with perturbation and as earlier $\hat{Y}_i$ represents a column vector of the data matrix.

Table 4. 1 Description of main sysmbols in chapter 4

| Symbol | Description |
|--------|-------------|
| $N_i$ | $i^{th}$ IoMT device |
| $Y$ | Global matrix |
| $Y_i(t)$ | The $i^{th}$ column of the global matrix |
| $\hat{\ }$ | Perturbed version of the original symbol |
| $\delta_i$ | The $i^{th}$ filter parameter |
| $\theta_i$ | Prediction model of $i^{th}$ IoMT data |
| $\ell$ | Mean square error function |
| $\alpha$ | Learning rate/step size |
| $\lambda$ | Eigen value of a matrix |
| $\Delta$ | The perturbation error |

The perturbation in the global data matrix is due to errors caused by filtering and predicton. The formation of aggregated data matrix is discussed in Section 4.5. The fog server's role is threefold. Firstly it estimates/predicts the perturbed data matrix ($\hat{Y}$), and secondly it computes and delivers filter parameter ($\delta_i$) for all the IoMT devices, and finally, it performs decision making using the perturbed data matrix. Table 4.1 shows some useful notations. In the beginning, all the IoMT nodes train the prediction model by running several instances of Least Mean Square (LMS) filter (section 4.4.1). Both the local IoMT device and the fog server uses the same prediction scheme. The local IoMT device runs a local processing subroutine as described in Algorithm 1 and the fog server runs Algorithm 2.

## 4.4    Theoretical Analysis

### 4.4.1  Adaptive Filtering in IoMT Devices

Adaptive filters are typically implemented for signals with non-stationary statistics and where no prior information is available. A typical adaptive filter is depicted in Fig. 4.2. Among various adaptive filters [49] this section selects Least Mean Square (LMS) filter [50] for local processing inside the IoMT node, since it has a very low computational overhead. Let for an IoMT device $N_i$ at time t the predicted IoMT sensor vector $\hat{Y}_i(t)$ be a linear approximation of the real sensor vector $Y_i(t)$. The LMS adaptive filter embedded inside the IoMT devices aims to minimise the error function $e(t)$, which is the least mean square approximation between the predicted sensor vector and the real sensor vector.

$$e_i(t) = \frac{1}{2}\sum_{i=1}^{n}\left(\hat{Y}_i(t) - Y_i(t)\right)^2 \tag{4.1}$$

The relationship between the predicted sensor vector $\hat{Y}_i(t)$ (output of LMS filter) and the real sensor vector $Y_i(t)$ is as follows.

$$\hat{Y}_i(t) = \theta_i^T Y_i(t) \tag{4.2}$$

The LMS filter relies on the stochastic gradient descent (SGD) optimisation, this approach takes iterative steps ($\alpha_i$) towards the steepest decrease of the error function $e_i(t)$. Eq. 4.3 shows the LMS update rule also known as Widrow-Hoff learning rule.

$$\theta_i(t) = \theta_i(t-1) + \alpha_i(t) \cdot e_i(t) \cdot Y_i(t) \tag{4.3}$$

Based on the empirical observation [49] to ensure convergency the step size $\alpha_i(t)$ should satisfy the following.

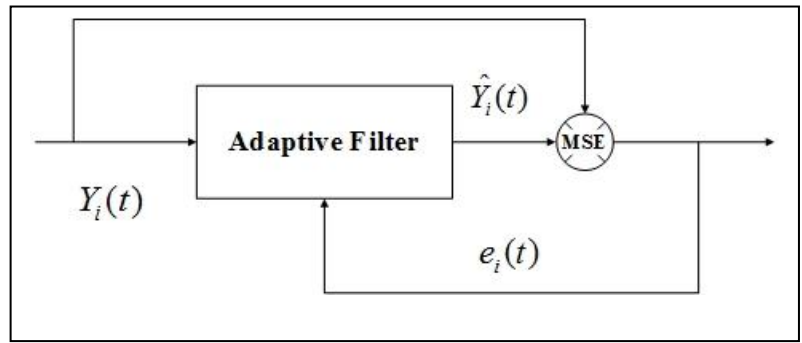

Fig. 4. 2 Block diagram of an adaptive filter

$$0 \leq \alpha_i(t) \leq \frac{1}{P_Y} \tag{4.4}$$

Where, $P_Y = \dfrac{1}{M} \sum_{j=1}^{M} |Y_i(j)|^2$, and M is the number of iterations taken for training the LMS filter.

### 4.4.2  Perturbation Analysis at Fog Server

The filter parameters play a key role in balancing the tradeoff between the desirable loss of decision accuracy (by allowing perturbation to $\hat{Y}$) and low communication overhead. This chapter uses the matrix perturbation theory [51] to bound the perturbation error ($\Delta$) of the perturbed data matrix which in turn affects the decision accuracy. The fog server generates a perturbed data matrix $\hat{Y} = Y + W$, where $W$ is the perturbation/filtering error and column elements of $W$, $W_i \in [-\delta_i, \delta_i]$. Let the $\lambda_i$ and $\hat{\lambda}_i$ denote the eigenvalues of the real covariance matrix $A = \dfrac{1}{m} Y^T Y$ and the perturbed covariance matrix

$\hat{A} = \dfrac{1}{m} \hat{Y}^T \hat{Y}$ respectively.

The norm of the perturbation error matrix $\Delta = A - \hat{A}$ can be formulated using the property of triangle inequality [52] is depicted as follows.

$$
\begin{aligned}
\|\Delta\| &= \left\| Y^T W + W^T Y + W^T W \right\| \\
&\leq \left\| Y^T W \right\| + \left\| W^T Y \right\| + \left\| W^T W \right\|
\end{aligned}
\tag{4.5}
$$

The goal here is to determine an upper bound for the expectation of RHS in the above inequality.

This chapter assumes that all the column vectors of $W$ are independent and all the column elements are i.i.d random variables with zero mean ($\mu = 0$) and variance $\sigma_i^2 \approx \sigma_i^2(\delta_i)$ along with fourth moment as $\mu_i^4 = \mu_i^4(\delta_i)$.

Using Jensen inequality $E(x) \leq \sqrt{E(x^2)}$.

$$
\begin{aligned}
E\left( \|\Delta\|_F \right) &\leq 2 E\left( \left\| Y^T W \right\|_F \right) + E\left( \left\| W^T W \right\|_F \right) \\
&\leq 2 \sqrt{E\left( \left\| Y^T W \right\|_F^2 \right)} + \sqrt{E\left( \left\| W^T W \right\|_F^2 \right)}
\end{aligned}
\tag{4.6}
$$

Based on Mirsky's theorem [51].

$$
E\left( \sqrt{\frac{1}{n} \sum_{i=1}^{n} \left( \hat{\lambda}_i - \lambda_i \right)^2} \right) \leq E\left( \frac{\|\Delta\|_F}{n} \right) \leq Tol_F
\tag{4.7}
$$

$$
E(\|\Delta\|_F) \leq 2 \sqrt{\frac{1}{m^2 n} Tr\left( Y^T Y \right) \cdot \sum_{i=1}^{n} \sigma_i^2} + \sqrt{\left( \frac{1}{m} + \frac{1}{n} \right) \cdot \sum_{i=1}^{n} \sigma_i^4}
\tag{4.8}
$$

$$
E(\|\Delta\|_F) \leq Tol_F
\tag{4.9}
$$

The Eq. 4.9 presents an upper bound ($Tol_F$) on the perturbation error caused due to local filtering at IoMT devices and estimation of perturbed data matrix using outdated shared model.

$$Tol_F = 2\sqrt{\frac{1}{m^2 n} Tr\left(Y^T Y\right) \cdot \sum_{i=1}^{n} \sigma_i^2} \;\; + \sqrt{\left(\frac{1}{m} + \frac{1}{n}\right) \cdot \sum_{i=1}^{n} \sigma_i^4} \tag{4.10}$$

Similar upper bounds can also be derived using spectral norm $\left\| . \right\|_2$, moreover this chapter selects Frobenious norm $\left\| . \right\|_F$ for no particular reason.

### 4.4.3  Uniform Filter Parameter Selection

This chapter assumes an independent and uniform distribution of IoT filter parameter within the interval $\left[-\delta_i, \delta_i\right]$. Moreover we also assume a homogeneous filter parameter allocation among all the IoMT devices, therefore $\delta_i = \delta$ and $\sigma_i = \dfrac{\delta^2}{3}$.   Solving Eq. 4.10 for $\delta$.

$$\delta = \frac{\sqrt{\dfrac{3Tr(Y^T Y)}{m} + 3 \cdot Tol_F \cdot \sqrt{nm + m^2}} - \sqrt{\dfrac{3 \cdot Tr(Y^T Y)}{m}}}{\sqrt{m+n}} \tag{4.11}$$

The Eq. 4.11 provides a relationship between local filtering and the global perturbation error, that plays a crucial role in balancing the tradeoff between local filtering at IoMT devices and the global eigen perturbation error.

## 4.5    Federated Filtering Framework

Federated Filtering Framework (FFF) is based on a loose federation of the participating devices (IoMT devices) those are coordinated by the central fog server. The FFF consists of two crucial protocols first, the local data processing protocol and the second is global data processing and coordination protocol.

### 4.5.1 Local Processing Protocol at IoT Device

Given the severe resource constraints in computation for IoMT devices, this section proposes a very lightweight filtering protocol for local processing. The local filtering is based on LMS adaptive filter (section 4.4.1). The IoMT devices computes a local prediction model $\theta_i$ (Eq. 4.3) from the collected data and share this model with the fog server. Now assuming $\theta_i$ as the current prediction model and $\delta_i$ as the latest filtering parameter for the $N_i$ IoMT device. $N_i$ at any time t tracks the deviation of predicted sensor vector $\hat{Y}_i(t)$ from real sensor vector $Y_i(t)$ using $W_i(t) = Y_i(t) - \hat{Y}_i(t)$. Whenever $|W_i(t)| > \delta_i$ the IoMT device updates the prediction model $\theta_i(t)$ and resets $W_i(t)$ to zero. The updated prediction model along with a small amount of sample data is shared with the fog server. However the LMS filter incurs negligible computational overhead that enables the IoMT device to run multiple instances of filtering for better accuracy. The above mentioned details for local processing at IoMT devices is summarized in algorithm 1.

---

**Algorithm 1：Local Processing Protocol**

---

**Input:** current $\theta_i(t), \delta_i(t), SVD(\overline{Y_0})$, $Y_i(t)$ and $\alpha_i(t)$

**Output：** $\theta_i^*(t)$

---

1: **for** (true) **do**
2:　　　t= current time
3:　　　compute: $W_i(t) = Y_i(t) - \hat{Y}_i(t)$
4：　　　**if** $|W_i(t)| > \delta_i$ **then**
5：　　　　$[\theta_i^*(t)] := LMS(Y_i(t), \alpha_i(t))$
6：　　　　$N_i$ sends (i, $\theta_i^*(t)$, $Y_i(t)$) to fog server
7：　　　　Set $W_i(t) \leftarrow 0$
8：　　　　Set $\theta_i(t) \leftarrow \theta_i^*(t)$
9：　　　**end if**
10:　**end for**

---

## 4.5.2  Federated Processing at Fog Server

At the beginning of each round the fog servers updates the current prediction models with the newly shared models. The fog server selects a random fraction K of the n participating IoMT devices. This chapter selects a random fraction of IoMT devices [47] since the decision accuracy degrades beyond a certain number. The step size $\alpha_i(t)$ is kept constant based on the empirical result (section 4.4.1). The fog server aggregates the model using Eq. 4.12.

$$\eta_i(t) = \sum_{i=1}^{K} \frac{n_k}{n} \theta_i(t-1) \tag{4.12}$$

Thereafter the fog server predicts the perturbed data matrix using the following equation.

$$\hat{Y}(t) = \eta_i^T Y(t) \tag{4.13}$$

The perturbed data matrix $\hat{Y}(t)$ is used for decision making. The impact of eigen perturbation error on the decision accuracy can be studied in [52]. The fog server continuously tracks $E(\|\Delta\|_F) > Tol_F$. Once the data matrix perturbation error exceeds the tolerable perturbation error threshold, the fog shares an updated filter parameter and summons all IoMT devices to share their updated prediction model. The above mentioned scheme is summarized as Algorithm 2.

**Advantages:** The proposed framework minimises the communication overhead (section 4.6) by limiting the number of transmissions to the central server. The algorithm 2, i.e. the model averaging makes it practically impossible to extract an individual model from the average model; that ensures privacy to sensitive medical data. Furthermore, the fog server, unlike a cloud server, is located closer to the source, which reduces the latency.

## 4.6    Performance Evaluation

In this section, we present some experimental results based on real IoT health data.

---

**Algorithm 2：Filter Model Averaging**

---

1: **for**   (true)   **do**
2:        t= current time
3:        **if**   $E(\left\|\Delta\right\|_F) \leq Tol_F$ **then**
4：       $\eta_i(t) \leftarrow \sum_{i=1}^{K} \dfrac{n_k}{n} \theta_i(t-1)$
5：       $\hat{Y}(t) \leftarrow \eta_i^T Y(t)$
6：       Perform decision making
7:        **else**
8:         Fog server shares  $\delta_i$  with  $N_i$
9:         Fog server receives (i,  $\theta_i^*(t)$,  $Y_i(t)$)
9:         **end if**
7:    **end for**

---

The results include the prediction using the filter model averaging by the fog server, the plot of communication overhead while varying local filtering parameter and the overall scalability of the proposal concerning energy efficiency. The experiments are performed using a publicly available real IoT health dataset known as MHEALTH data (available at: http://archive.ics.uci.edu/ml/datasets/MHEALTH+Dataset). The dataset comprises body motion and vital signs recordings for ten volunteers of diverse profile while performing 12 physical activities. For our experiments, we have only considered the chest accelerometer sensor reading, i.e. columns 1-3 and the right lower arm gyro sensor time series data, i.e. column 18-20.

Based on section 4.4.3 we assume a homogeneous filter parameter for all the IoMT devices. We initially distribute the data equally among 50 IoMT devices and compute normalized tolerable perturbation error as shown in Eq. 4.14.

$$\left\langle Tol_F \right\rangle = Tol_F \Big/ \sqrt{\frac{\sum \lambda_i^2}{n}} \qquad (4.14)$$

We present the relationship between the normalized tolerable perturbation error and the uniform filter parameter in Fig. 4.3. Moreover Fig. 4.3 depicts a roughly linear relationship between the normalized tolerable perturbation error and the local filter parameter. It is also intuitive since whenever one increases the $\langle Tol_F \rangle$, the filter at IoMT devices passes more data.

Next, we present the prediction performance of the filter model averaging scheme (Algorithm 2) by the fog server. Due to space limitations, we offer prediction results of two different IoMT devices (Fig. 4.4).   As discussed in section 4.5 both the local and the global filtering uses the same LMS filter. The available sophisticated techniques that provide better accuracy cannot be used at the fog server since those techniques must also be feasible for local processing at IoMT devices. Given the severe resource constraints in power and computation, the sophisticated methods cannot be used by IoMT devices for local processing.

Towards this end, we plot the communication overhead as a function of filter parameter. We observe that in Fig. 4.5 the communication cost can be massively reduced even with a tolerable perturbation error. Based on the performance evaluation, we have achieved upto 95 % reduction in transmissions as shown in Fig. 4.6 (at $\delta = 2.0$). This supports our claim that the proposed framework can provide a good tradeoff between communication efficiency and eigen perturbation error of data matrix.

Finally, we examine the scalability of the proposed scheme for small as well as a large number of devices. The energy efficiency ($\eta$) of the system with n number of IoMT devices can be computed as:

$$\eta = \sum_n \frac{d_n}{E_n \cdot r_n \cdot TTI} \qquad (4.15)$$

Where  $d_n$  is the total volume of data to be uploaded,  $E_n$  is the average energy consumed to deliver a single packet,  $r_n$  is total number of data packets to be uploaded by all the IoMT devices and TTI = 1 is transmission time interval which is constant to all packets. It is evident from the plot (Fig. 4.6) that our FFF scheme is highly scalable compared to other recent researches such as AM-DR [46] and well known ARIMA [45]. Based on the plot, the energy efficiency increases with the number of devices. Therefore the proposed framework can also be extended to a massive IoMT scenario.

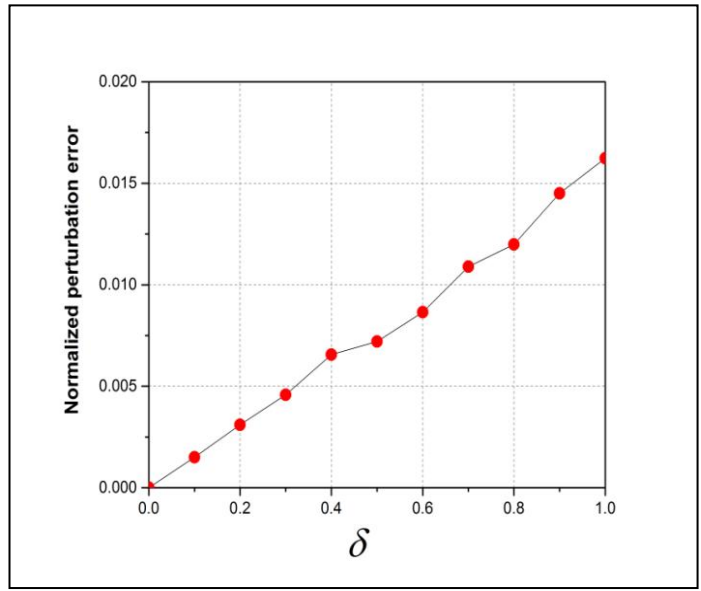

Fig. 4. 3 Normalized Tolerable Perturbation Error as a function of $\eta$

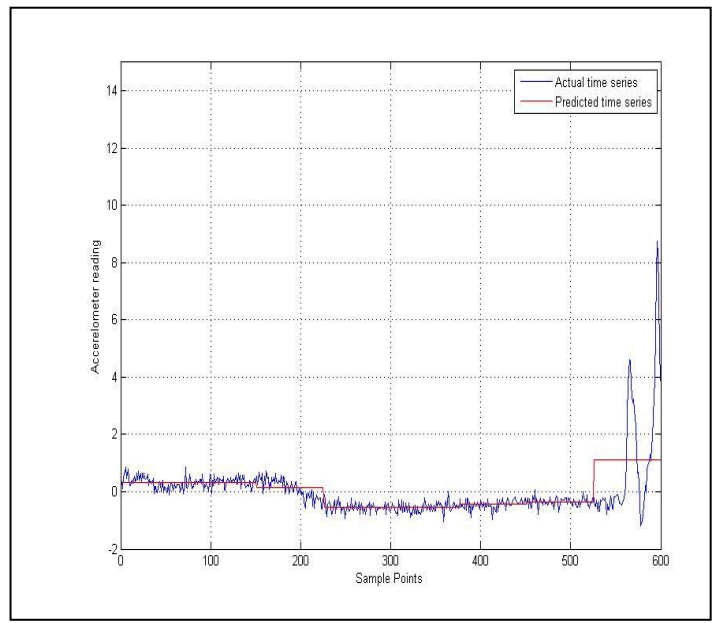

Fig. 4. 4 Prediction performance of accelerometer sensor readings

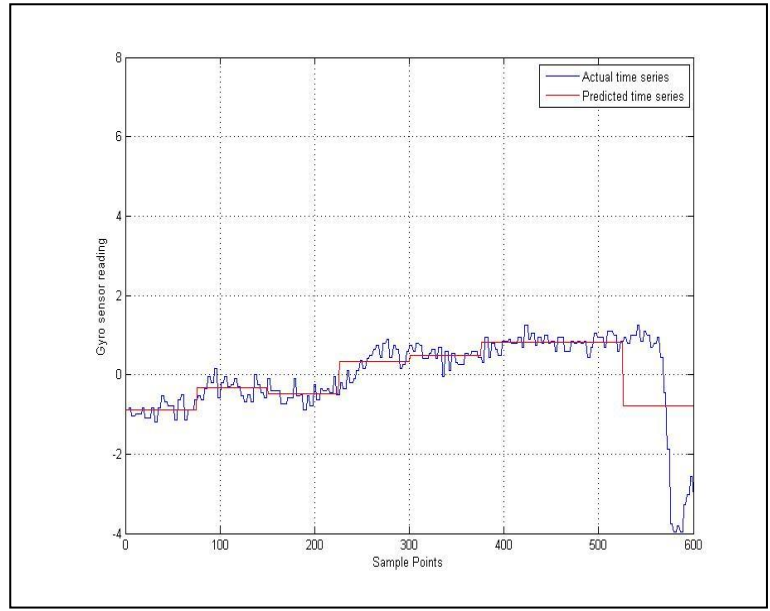

Fig. 4. 5 Prediction performance of gyro sensor readings

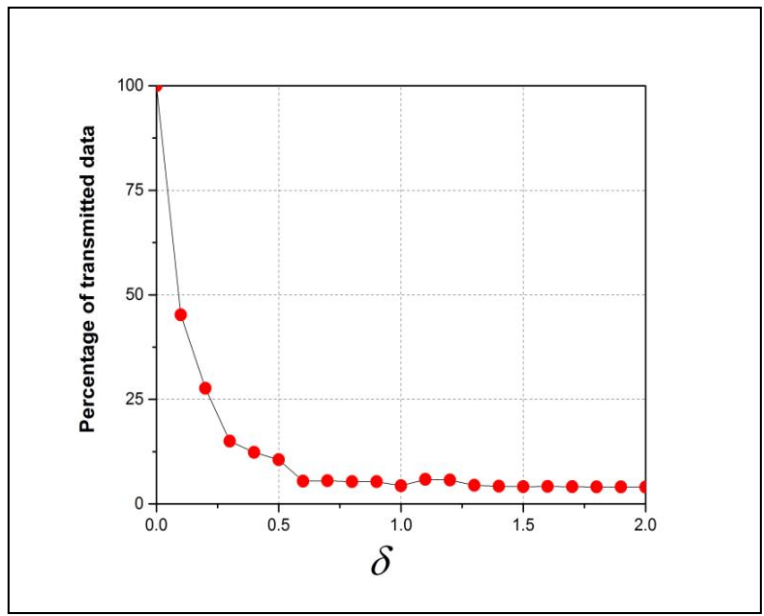

Fig. 4. 6 Communication overhead as a function of $\delta$

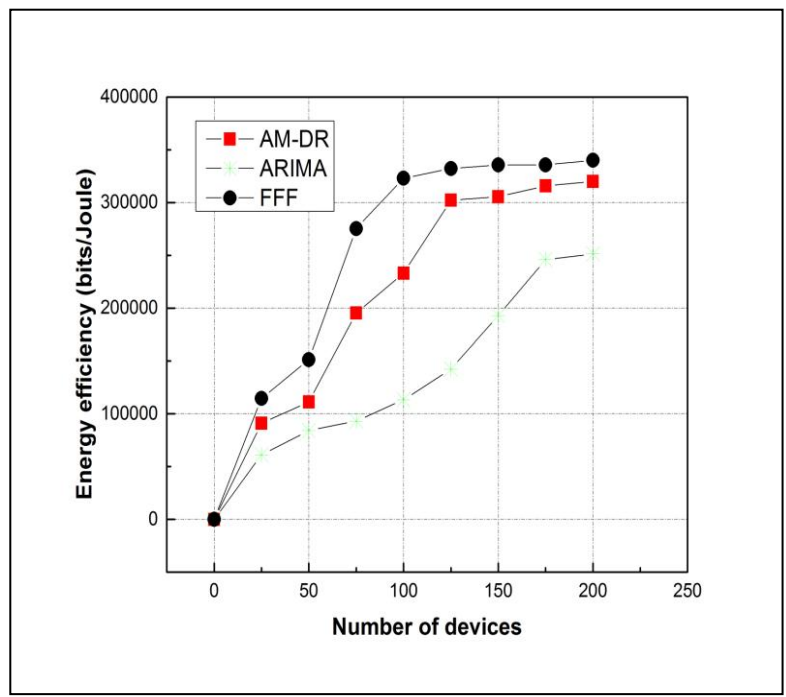

Fig. 4. 7 Energy efficiency as a function of number of devices

## 4.7    Chapter Conclusion

This chapter considers open challenges concerning energy efficiency, privacy and latency for smart healthcare analytics. This chapter derives a theoretical upper bound on the eigenvalue perturbation and further formulates a relationship between the local quantization at IoMT devices with the global perturbation error at fog server. Based on the theoretical infrastructure this chapter proposes two subroutines first for the local filtering at the IoMT device and the second for the central fog server. The proposed framework cuts down 95 % of the communication overhead. Moreover, the use of perturbed data matrix (predicted data) instead of using real global matrix for decision making ensures better privacy and the low proximity of fog server provides low latency.

.

# Chapter 5 Conclusion and Futurework

## 5.1    Conclusion

The IoT is experiencing an unprecedented growth regarding the number of interconnected devices, wireless sensors and amount of generated data. The data generated by the massive IoT networks if analyzed properly can provide several actionable insights to facilitate decisionmaking at industries and societies. Therefore, IoT analytics is widely acknowledged as one of the key enablers of a new industrial revolution (Industry 4.0). However, the dominant IoT analytics architecture that performs excellent historical IoT analytics does not serve well in the case of proactive IoT analytics. This dissertation envisions three key open challenges: a) resource constraints of IoT devices concerning battery power, computational capability and storage; b) the problem of veracity in highly uncertain IoT data; c) network latency and privacy issues for critical healthcare applications. This dissertation proposed a four-layer data analytics architecture aims to solve the open challenges using data aggregation (routing and processing) prior to IoT analytics. Overall this dissertation solves one or more problems by proposing fully unsupervised data-driven algorithmic and theoretical frameworks for various IoT related scenarios and presents simulation studies to establish the efficacy of the proposals.

Routing is a major part of data aggregation and providing an optimized and energy efficient routing scheme is one of the major open challenges for analytics systems. Moreover, it is observed that the literature lacks a unified approach to IoT data routing that supports both stationary and nonstationary IoT devices. Chapter 2 of this dissertation proposes a mobility based clustering approach. Later this chapter discussed a D2D communication based intercluster multi-hop data delivery scheme that supports a massive number of stationary and non-stationary IoT devices.

The IoT data is highly uncertain due to the presence of outliers, noise, missing values and redundancy. Therefore the unprocessed IoT sensor data is not fit to be used by analytics systems. This problem is considered in chapter 3. This chapter proposes a data aggregation scheme for highly uncertain raw IoT sensor data collected using the D2D communication. The approach initially reconstructs the subspace using sample data, and then it iteratively tracks down the low-rank approximation of the dominant subspace in the presence of high uncertainties at the fog server. Later, the robust dominant subspace is used to estimate a more reliable true sensor data matrix from the highly uncertain raw IoT sensor data traffic matrix. Moreover, the proposed scheme requires no prior information and is fully unsupervised.

The IoT analytics based on cloud computing typically pushes all the data to the remote cloud servers for analytics that causes high communication overhead, network latency and privacy issues. Chapter 3 considers the scenario of medical IoT analytics using massive IoMT devices. This paper presents a novel Federated Filtering Framework for IoMT devices which is based on the prediction of data at the central fog server using shared models provided by the local IoMT devices. The fog server performs model averaging to predict the aggregated data matrix and also computes filter parameters for local IoMT devices. Theoretical upper bounds on data perturbation and a consensus between distributed local filtering and global data perturbation are key theoretical contributions of this chapter.

## 5.2    Future work

This dissertation has demonstrated that data aggregation is a critical component in the IoT analytics architecture and can optimize and control the performance of the analytics system. In order to enable data aggregation to support a broader range of applications in large scale IoT analytics systems, there are several exciting extensions to our work. The multi-hop inter-cluster data routing in chapter 2 can be further extended by taking consideration of more parameters such as available battery power and channels for data transmission. Moreover, the scenario can be formulated into an optimization problem with constraints depending on the application scenario. In Chapter 4 the federated filtering framework can be extended in two significant ways; first, by using a very sophisticated compressed machine learning algorithm such as decision trees for local filtering that could have provided better superior prediction accuracy; second, by formulating an optimization problem for the filter parameters instead of assuming it to be homogeneous.

# Acknowledgement

There are many people whom I wish to thank for their direct and indirect contributions in making this dissertation possible.

First and foremost, I want to thank my parents, Mita Sanyal and Asit Bandhu Sanyal, for supporting my dreams and ambitions in every possible way. Their prayers and encouragements have helped me to get through the physical, emotional and intellectual tidal waves of the graduate studies.

Next, I would like to thank my advisor, Prof. Dapeng Wu, for the guidance, support, and honest criticisms he has provided throughout this work. I am grateful for the numerous meetings that he had with me to discuss our research problems and to guide me through them with his profound knowledge in many diverse areas. His mastery in the research domain has saved me on more than one occasion. Without his invaluable support and constant guidance, I could not have completed this dissertation.

I would also like to thank my external research collaborators: Boubakr Nour, Animesh Chattopadhyay and internal research collaborators: Prof. Puning Zhang, Prof. Pingping Zhang, Junjie Yan, Bin Zhou, Jean Bosco. They all have dedicated several hours on ideation, reviewing and discussing theoretical work for the dissertation. I have gained many insights on academic research while working with them on various research projects.

Finally, I would like to thank my teachers from the coursework: Prof. Rong Chai, Prof. Ning-hai Bao, Prof. Xiaoge Huang and my friends: Faith Birnstein, Subhadeep, Ananya, Waleed, Abubakar, Mostofa, Ahmad, Nosheen and Yaqoob. On the one hand, my tutors have forged my character and scientific foundation through their lectures. On the other hand, my friends have helped me to cope up with a new academic and living environment. Most importantly I am very much grateful to the Chinese government scholarship council for the generous scholarship to support my education in Chongqing, China.

# Publications and Achievements

## Publications

- **Sunny Sanyal**, Dapeng Wu, Boubakr Nour, "A Federated Filtering Framework for Internet of Medical Things", IEEE ICC 2019 conference, Shanghai, China, 20-24 May 2019.  (Presented)

- **Sunny Sanyal**, Bin Zhou, Animesh, Jean Bosco, "DeepMines: A fog enabled prediction platform for underground coal mines", Project Report, IEEE ComSoc Student's Competition 2018.  **[IEEE Comsoc honorary mention award]**

- **Sunny Sanyal**, Puning Zhang, "Improving Quality of Data: IoT data aggregation using device to device communications", IEEE Access, Vol. 6, page. 1-11, Nov. 2018.

- **Sunny Sanyal**, Wu Dapeng, Junjie Yan, Xing Li,"Co-relative mobility based IoT data uploading using D2D communication", In Proc. of 10th EAI International Conference on Mobile Multimedia Communications, Chongqing, China, 13-14 July 2017.

- Dapeng Wu, Junjie Yan, **Sunny Sanyal**, Ruyan Wang, "Trust oriented partner selection in D2D cooperative communications", IEEE Access, Vol. 5, page. 3444-3453, February 2017.

## Achievements

- Honorary Mention Award in IEEE ComSoc Student's Competition 2018.

- Excellent Master's Thesis Award, Chongqing University of Posts and Telecommunications (including all departments).

- Outstanding International Student Award, International College 2019, CQUPT, China.

- CQUPT Technology Innovation Award 2019 (科技创新奖), International College, CQUPT, China.

- Chinese Government Scholarship with a round trip travel grant.

- IEEE Accommodation grant for IEEE ComSoc summer school 2018.

- ▪ Services towards academic journal and conferences: IEEE Access, IEEE VTC 2019, IEEE ICC 2019, IEEE ICCCN 2019, IEEE ICEET 2019, ACM SAC 2019, EAI FABULOUS 2019, EAI UBICNET 2019 and EAI INTERSOL 2019 in different roles such as technical program committee, program committee, publicity co-chair and reviewer.